\documentclass[useAMS,usenatbib,letter]{mnras}
 
\usepackage{mathptmx}
\usepackage{lscape}
\usepackage{rotating}
\usepackage{amssymb}
\usepackage{xspace}
\usepackage{color}
\usepackage{subfigure}
\usepackage{bm}
\usepackage{graphicx}
\usepackage[T1]{fontenc}
\def\simgt{\lower 2pt \hbox{$\, \buildrel {\scriptstyle >}\over {\scriptstyle \sim}\,$}}
\def\simlt{\lower 2pt \hbox{$\, \buildrel {\scriptstyle <}\over {\scriptstyle \sim}\,$}}
\def\chandra{{\it Chandra}}

\def\xmm{\hbox{\it XMM-Newton}}

\def\xray{\hbox{X-ray}}

\def\arcsec{^{\prime\prime}}
\def\lx{$L_{\rm X}$}
\def\lr{$L_{\rm radio}$}
\def\hz{high-$z$}

\def\tnj{TN~J1338$-$1942}

\begin{document}

% -----------------------------------------------------------------------------
% Title
% -----------------------------------------------------------------------------

\title[CMB radio quenching of jetted AGNs]{CMB-induced radio quenching of high-redshift jetted AGNs with highly magnetic hotspots}

% -----------------------------------------------------------------------------
% Authors and their affiliations
% -----------------------------------------------------------------------------

%\author{\rm Jianfeng~Wu, Jerome~A.~Orosz, Jeffrey~E.~McClintock,
%  Lijun~Gou, Imran~Hasan, Charles~D.~Bailyn}

\author[Jianfeng Wu et al.]
{Jianfeng~Wu$^1$\thanks{Email: jfwu@umich.edu},
Gabriele~Ghisellini$^2$,
Edmund~Hodges-Kluck$^1$,
Elena~Gallo$^1$,
\newauthor Benedetta~Ciardi$^3$,
Francesco~Haardt$^{4,5}$,
Tullia~Sbarrato$^6$,
Fabrizio~Tavecchio$^2$
  \\
  $^1$ Department of Astronomy, University of Michigan, 1085 S University Ave, Ann Arbor, MI 48109, USA \\
  $^2$ INAF -- Osservatorio Astronomico di Brera, Via Bianchi 46, I--23807 Merate, Italy\\
  $^3$ Max Planck Institute for Astrophysics, Karl-Schwarzschild Strasse 1, D-85741 Garching, Germany\\
  $^4$ DiSAT, Universit\`{a} dell'Insubria, via Valleggio 11, I-22100 Como, Italy\\
  $^5$ INFN, Sezione di Milano-Bicocca, Piazza della Scienza 3, I-20126 Milano, Italy\\
  $^6$ Dipartimento di Fisica `G. Occhialini', Universit\`{a} di Milano-Bicocca, P.za della Scienza 3, I-20126 Milano, Italy  
  }

% -----------------------------------------------------------------------------
% Abstract
% -----------------------------------------------------------------------------

\maketitle
\begin{abstract}

In an effort to understand the cause of the apparent depletion in the number density of radio-loud AGNs at $z>3$, this 
work investigates the viability of the so-called Cosmic Microwave Background (CMB) quenching mechanism of intrinsically jetted, high-$z$ AGNs, whereby Inverse Compton scattering of CMB photons off electrons within the extended lobes results in a substantial dimming of the lobe synchrotron emission at GHz frequencies, while simultaneously boosting their diffuse \xray\ signal. 
We focus on five $z>3.5$ radio galaxies that have
sufficiently deep \chandra\ exposure (> 50~ks) to warrant a meaningful investigation of any extended \xray\ emission. For those objects with evidence for statistically significant extended \xray\ lobes  (4C~41.17 and 4C~03.24), we combine the \chandra\ measurements with literature data at lower frequencies to assemble the systems' Spectral Energy Distributions (SEDs), and utilize state-of-the-art SED modelling \citep{ghisellini+2015}  -- including emission from the disk, torus, jet, hotspots, and lobes -- to infer their physical parameters. 
For both radio galaxies, the magnetic energy density in the hotspots is found to exceed the energy density in CMB photons, wheres the opposite is true for the lobes. This implies that any extended synchrotron emission likely originates from the hotspots themselves, rather than the lobes. Conversely, Inverse Compton scattering of CMB photons dominates the extended \xray\ emission from the lobes, which are effectively "radio-quenched". As a result, CMB quenching is effective in these systems in spite of the fact that the observed \xray\ to radio luminosity ratio does not bear the signature $(1+z)^4$ dependence of the CMB energy density.  

\end{abstract}


% -----------------------------------------------------------------------------
% Introduction
% -----------------------------------------------------------------------------

\section{Introduction}\label{intro}

Radio synchrotron emission from jetted Active Galactic Nuclei (AGNs) arises from magnetized plasma within collimated, relativistic jets, plus the jet-inflated lobes, including the hotspots. If the observer line of sight lies within an angle $\sim 1/\Gamma$ of the jet axis, where $\Gamma$ is the jet bulk Lorentz factor, then the beamed jet emission dominates the energy spectrum, and the system is classified as a blazar ($\Gamma\simeq 10-15$ for typical blazars). For large viewing angles, isotropic emission from the extended lobes takes over, and a powerful radio galaxy is observed. From simple beaming arguments, it follows that, for each observed blazar, there ought to be $\sim 2\Gamma^2$ misaligned, jetted AGNs. 
The identification of more than a dozen high-redshift blazars ($z\simgt4$)
\citep[e.g.,][and the references therein]{yuan+2006,sbarrato+2013,
sbarrato+2015,wu+2013} implies the existence of a much larger population of jetted AGNs only $\simlt1.5$~Gyr after the Big Bang. Nevertheless, in spite of being well within the detection threshold of modern wide-field radio surveys (such as the FIRST survey\footnote{Faint Images of the Radio Sky at
Twenty Centimeters \citep{becker+1995}.}), this high-$z$ ``parent population'' remains elusive. 
\citet{volonteri+2011} find that the expected number density of radio-loud quasars, as inferred from the the hard \xray\ detected, BAT\footnote{Burst Alert Telescope, onboard the {\it Swift} satellite \citep{krimm+2013}.} sample of luminous, massive blazars \citep{ajello+2009}, significantly overestimates the number of observed luminous, {\it radio-loud quasars} detected by the SDSS\footnote{Sloan Digital Sky Survey \citep{york+2000}.} at $z\simgt 3$. Qualitatively similar
conclusions are reached by \cite{ghisellini+2016}, \citet{kratzer+2015} (see also \citealt{haiman+2004,mcgreer+2009}). 

Possible explanations for this apparent deficit include \citep[e.g.,][]{volonteri+2011}: selection biases, such as heavy optical obscuration by dust; systematic, intrinsic differences 
in the jets' physical properties at high-$z$ compared to the local population (such as lower
average $\Gamma$ factor); and/or intrinsic dimming of the radio lobes. 
The first hypothesis implies the existence of a large population of
infrared-luminous, radio-loud quasars with no detectable optical counterparts,
whereas the second appears at odds with the notion of the most powerful
jets being associated with high accretion rates, likely typical at early cosmic times. 

With respect to the third hypothesis, an attractive possibility is that of Cosmic Microwave Background (CMB) photons affecting the behavior of jetted AGNs \citep[e.g.,][]{celotti+2004,mocz+2011}. 
In a series of recent works, Ghisellini et al. \citep{ghisellini+2014,ghisellini+2015,ghisellini+2016} explore specifically how the interaction
between the CMB radiation and electrons within jet-powered lobes affects the observational appearance of jetted AGNs at different redshifts. 
Simply put, owing to its $(1+z)^4$ dependence, the CMB energy density likely starts to dominate over magnetic energy density within the lobes above $z\simeq 3$. 
As a result, synchrotron radio emission is progressively suppressed at higher redshifts, while the high-energy electrons cool effectively by Inverse Compton scattering off of CMB photons (IC/CMB). The cooling timescale of the IC/CMB
mechanism at high redshifts is much shorter than typical jet lifetime (see the formulation in Section 2.2 of \citealt{ghisellini+2014}). For example, at $z\simgt3$, for electron Lorentz factor of $\gamma\sim 4\times 10^3$, the cooling time $\gamma/\dot{\gamma}\sim 2$~Myr. This leads to dimming in the radio and enhancement in the X-rays. Even if the CMB energy does not exceed the typical jet or hot spot magnetic energy density one nonetheless expects radio dimming and \xray\ enhancement in these features as well, and IC emission is seen in \hz\ \xray\ jets \citep[e.g.,][]{siemiginowska+2003,cheung+2006,cheung+2012,mckeough+2016}. Systems where IC/CMB cooling is dominant would then be classified as radio-quiet despite being jetted AGNs. 

Even though most \hz\ radio-loud quasars in the FIRST survey are ``core dominated'' in maps with a 5~arcsec beam size, this does not mean that all quasars with powerful jets are dominated by a flat-spectrum core where the IC/CMB mechanism would not work. Indeed, many of the core-dominated systems are steep-spectrum radio quasars (SSRQs), which are associated with misaligned jets and are evidently not dominated by the flat-spectrum core associated with blazars and the pc- or kpc-scale jets in local radio galaxies. Many of the core-dominated systems may be small radio galaxies analogous to low-$z$ compact-steep-spectrum sources, since at $1<z<4$ the FIRST beam corresponds to 35--40~kpc. From a physical standpoint, powerful jetted AGN share the basic features of a jet, a terminal shock at the interface of the jet head and surroundings, and spent jet material downstream that is associated with the radio lobes. Thus, the enhanced IC scattering could be relevant for any SSRQ, and we expect that the SSRQs comprise most of the ``missing'' sources, since flat-spectrum radio quasars (FSRQs) are identified with blazars.

This paper is concerned with the IC/CMB hypothesis for ``missing'' radio-loud quasars. Although the CMB energy density dependence with redshift is well established, directly testing this hypothesis requires measuring magnetic energy densities, which is challenging even for nearby systems. An alternate approach is to search for radio-quiet quasars with \xray\ jets or lobes. A census of hard \xray\ jets and lobes in radio-quiet quasars would show how much of the discrepancy between blazars and radio-loud quasars the IC/CMB hypothesis can explain. However, this is observationally challenging because typical SSRQ radio-lobe luminosities, cast to the hard X-rays at $z>3$, require deep (100\,ks) \textit{Chandra} exposures to be detected, and they could also be confused with diffuse \xray\ emission or be impossible to see against the backdrop of the intrinsic AGN \xray\ emission. In addition, there is no \textit{a priori} knowledge of which radio-quiet quasars to target. \xmm\ has an angular resolution that is twelve times worse than \textit{Chandra}, and would only be sensitive to large lobes in addition to the other problems. Moreover, as exemplified by the results from deep \xmm\ imaging observations of two of the highest-redshift radio-quiet quasars (ULAS~J1120$+$0641, at $z=7.1$, and SDSS~J1030$+$0524, at $z=6.3$), even in the absence of extended \xray\ emission, the resulting upper limits could still be reconciled with the presence of extremely powerful -- and yet undetected -- jets (\citealt{fabian+2014}, \citealt{mocz+2013}). 

  Another approach is to estimate the typical dimming based on \xray\ emission from known radio systems, and then extrapolate to estimate how many systems would be undetected in the radio. To first approximation, if CMB-quenching of \hz\ radio lobes is effective, one would expect the ratio between \xray\ luminosity and radio luminosity \lx/\lr\ to increase with $z$, eventually leading to the disappearance of most radio lobes. \citet{ghisellini+2015} show that {\it none} of the 13 known $z>4$ blazars show evidence for extended radio emission in their GHz spectra, which is inferred from the lack of spectral softening below the 1~GHz rest-frame emission and supports CMB-quenching. Meanwhile, \citet{wu+2013} find that $z\simgt4$ blazars exhibit a modest \xray\ enhancement by a factor of $\sim3$ compared to lower redshift counterparts with similar radio luminosity, optical/UV luminosity, and radio loudness. Simply based on luminosity ratios, this supports the scenario of CMB photon upscattering, but only on a fractional basis, i.e., $\sim 6$\% of the \xray\ emission at $z\sim 1$ is from IC/CMB, and this fraction increases to $\sim 70$\% at $z\sim 4$. This supports the increasing importance of IC/CMB cooling at \hz\ but not to the same extent. On the other hand, \citet{smail+2012} find no evidence for a $(1+z)^4$ increase in \lx/\lr\ for a sample of 10 powerful radio galaxies between $1.8 < z < 3.8$. One possible explanation for the lower-than-expected \lx/\lr\ ratio is that most of the seed photons for IC scattering in these systems are locally produced, far-IR photons from star formation as opposed to CMB photons (\citealt{smail+2013}). Another possible explanation is that the radio and \xray\ emission mostly come from separate regions in the system due to spatial variation in the ratio of photon and magnetic energy densities, in which case the total \lx/\lr\ is not meaningful. For example, the jet or hot spots may have a particularly high magnetic energy density, whereas the dimming could work as expected in the lobes. 

  One way to avoid this issue is to decompose the radio and \xray\ emission through high fidelity, high resolution images. However, in general the available multiwavelength data have a range of beam sizes, many of which are large. This leads to some ambiguity when identifying radio and \xray\ features. In addition, emission from jet knots is complex and can come from a range of physical conditions even in high resolution images, and it is unclear how to extrapolate to undetected jets (especially without knowing the viewing angle). Another issue is that the \xray\ photons in the \chandra\ band typically come from lower-energy electrons below the synchrotron peak, even at $z>3$, whereas the radio fluxes are typically above the synchrotron peak. Connecting the two requires some model if the peak is not in the covered region. These issues motivate the approach adopted in this paper, which is to construct self-consistent models of jetted AGN based on \citet{ghisellini+2015} using a broad spectral energy distribution (SED) to determine the level of \xray\ enhancement in the lobes. This allows for a comparison of the diffuse \xray\ and radio emission in known radio galaxies. Our goal is to determine whether the lobe emission is consistent with the IC/CMB behavior (as opposed to the case where the seed photons are from local star formation), in which case CMB-quenching could turn sources where the luminosity is dominated by the diffuse component into radio-quiet quasars.

We focus here on the handful of \hz\ ($z\simgt3.5$) radio-loud AGNs with deep \chandra\ imaging observations. We establish robust, conservative criteria to isolate a subset of objects with statistically significant, extended \xray\ emission (\S2). For those, we decompose the \xray\ emission into core and extended (lobes+hotspots) components, and model the resulting broadband SEDs following \citet{ghisellini+2015}, accounting for both CMB as well as IR emission from the host galaxy as possible Comptonization seed photons (\S3). We summarize and discuss our results in \S4. 

% Table~1: X-ray Observation Log
\begin{table*}
\centering
\begin{tabular}{lcccrccc}
%\tablecaption{High-$z$ Radio Galaxies and Their \chandra\ Observations\label{log_table}}
  %\tablewidth{0pt}
  \hline
  \hline
{Name} &  {R.A.} & {Dec.} & {$z$} & Obs.~ID & Date & Exposure & References \\
{} & {(J2000)} & {(J2000)} & {} & {} & {} & {Time (ks)} &{} \\
%\startdata
\hline
4C~03.24 & 12:45:38.38 & $+$03:23:21.5 & $3.57$ & 12288 & 2010-12-05 & 90.6 & 1 \\
4C~19.71 & 21:44:07.51 & $+$19:29:15.1 & $3.59$ & 12287 & 2010-08-23 & 36.0 & 1 \\
& & & & 13024 & 2010-08-26 & 53.7 &  \\
4C~41.17 & 06:50:52.18 & $+$41:30:31.2 & $3.79$ & 3208 & 2002-09-25 & 63.2 & 2 \\
& & & & 4379 & 2002-09-26 & 73.7 & \\
4C~60.07 & 05:12:55.10 & $+$60:30:51.1 & $3.79$ & 10489 & 2008-12-10 & 98.7 & 3 \\
TN~J1338$-$1942 & 13:38:26.11 & $-$19:42:31.3 & $4.11$ & 5735 & 2005-08-29 & 30.4 & 4\\ 
& & & & 6367 & 2005-08-31 & 21.3 & \\
& & & & 6368 & 2005-09-03 & 21.5 & \\
\hline
\hline
\end{tabular}
\vskip 0.4 true cm
\caption{
  High-$z$ Radio Galaxies and Their \chandra\ Observations.
  References: (1) \citet{smail+2012}; (2) \citet{scharf+2003}; (3) \citet{smail+2009}; (4) \citet{smail+2013}. The coordinates are of the point \xray\ source. The exposure time has been corrected for background flares and detector dead time.
  }
\label{log_table}
\end{table*}

% -----------------------------------------------------------------------------
% Data Analysis
% -----------------------------------------------------------------------------

\section{X-ray Data Reduction and Analysis}\label{data}

To allow a meaningful investigation of their extended emission, we focus 
on \hz\ radio galaxies -- here defined as having $z>3.5$ -- with archival \chandra\
observations with total exposures exceeding 50~ks. This exposure time limit, although admittedly
arbitrary, is necessary since the surface brightness of lobe \xray\ emission is usually low
\citep{croston+2005}. The sample is composed of five objects, whose observation logs are listed in Table~\ref{log_table}.

All the five radio galaxies were observed with the Advanced CCD Imaging
Spectrometer \citep[ACIS;][]{garmire+2003}. Standard {\sc ciao}
v4.7 scripts/procedures were employed to reduce the data. For each individual observation, we
reprocessed the archival data and generated new level 2 event and bad pixel
files using the script \verb+chandra_repro+ with the latest Calibration
database applied (CALDB v4.6.8). The algorithm {\sc edser} \citep[Energy-Dependent
  Subpixel Event Repositioning;][]{li+2004} was applied as default,
for the purpose of generating subpixel \xray\ images (see below). 
The absolute astrometry was then corrected by matching
an initial point source list, generated with the \verb+wavdetect+
algorithm \citep{freeman+2002}, to the source catalog of the Two Micron All Sky
Survey \citep[2MASS;][]{skrutskie+2006}, or the Sloan Digital Sky Survey Data
Release 9 \citep[SDSS DR9;][]{ahn+2012}. The typical positional error after the
astrometric correction is between $0.2^{\prime\prime}$ (for 4C 41.17 which has higher
\chandra\ counts) and $0.5^{\prime\prime}$ (for the other four objects). 

If present, background flares ($3\sigma$ above the average) were removed using the
{\sc ciao} script \verb+deflare+ (filtered time intervals were minimal
($<15\%$) for all observations). 

X-ray images were generated in the observer-frame soft 
(0.5--2.0~keV), hard (2.0--8.0~keV), and full (0.5--8.0) bands using events
with ASCA grades 0,2,3,4,6. Exposure maps and PSFs were also produced for the full-band images. Source detection was then carried out again on the full-band images using \verb+wavdetect+, with
a detection threshold of $10^{-6}$ and wavelet scales of 1, $\sqrt{2}$, 2,
$\sqrt{4}$, and 4 pixels. For objects with multiple epochs (4C~19.71,
4C~41.17, and \tnj), we merged the observations by reprojecting the level
2 event files with shorter exposure times to the one with the longest exposure
time. With the merged event file, we repeated the above procedure for generating \xray\
images, exposure maps, PSF, and point source lists. 
Figs.~\ref{4c41_fig}--\ref{other_fig} show the full-band \chandra\ images for
all the five radio galaxies. 

Next, we aim to establish a quantitative criterion to determine whether the
detected \xray\ emission is point-like vs. extended, and whether, in the
presence of extended emission, its location and spatial extent are consistent
with the position of the galaxies' radio lobes. The latter is typically done
by comparing the source's measured photon density profile to that of the PSF.
The results of this procedure are presented in the following subsections.

\subsection{4C 41.17}\label{data:4c41}

\begin{figure*}
    \centering
    \includegraphics[width=6.5in]{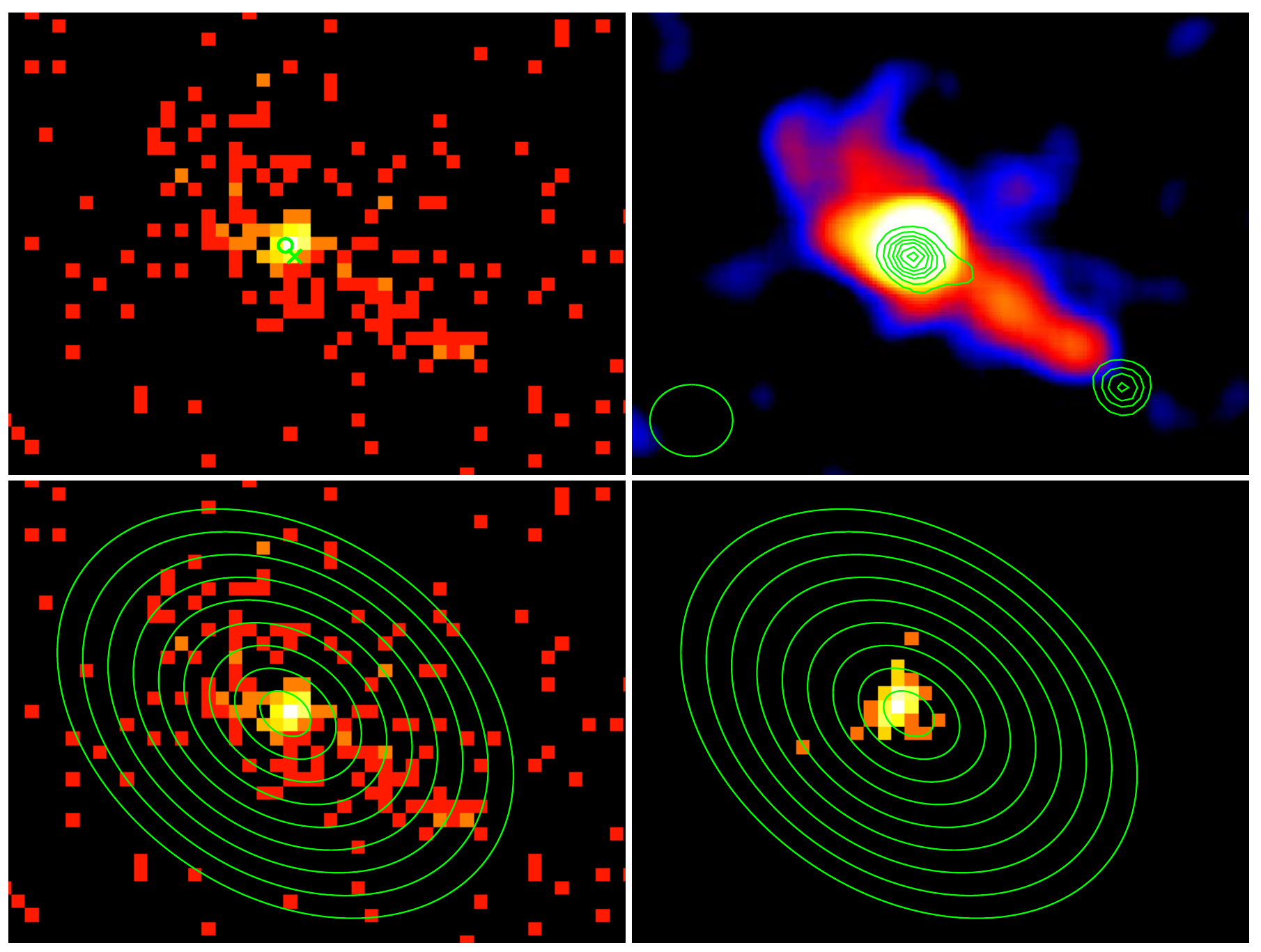}
    \caption{
      The \chandra\ ACIS images of 4C~41.17. The left two
        panels are the original full-band (0.5--8.0~keV) images. The upper right
        panel is the smoothed subpixel images, with the overlaid 1.4~GHz radio flux
        contours. The first level of the contours has the flux of 15mJy/beam, while each
        subsequent level has a flux increase of 15mJy/beam. The beam size is shown as
        the ellipse in the bottom left corner of this panel, with semi-major axis of
        $1.5^{\prime\prime}$ and semi-minor axis of $1.3^{\prime\prime}$.
        The lower right panel is the ChaRT
        simulated PSF for the \chandra\ observations. In the upper left panel, the
        open circle and the cross symbol label the \xray\ position and the radio position
        of the point source, respectively. The elliptical annular regions in the lower
        panels are those used in calculating the profile of the \xray\ count number
        density for the extended \xray\ emission.
    }
    \label{4c41_fig}
\end{figure*}

4C~41.17 has been observed twice by \chandra, yielding a total (good) exposure time of
137~ks. The upper left panel of Fig.~\ref{4c41_fig} shows the merged, full-band
\chandra\ image, where  the green open circle labels the position of the point
\xray\ source as identified by \verb+wavdetect+.
To inspect the source spatial profile, we first created a sub-pixel
full-band \xray\ image for 4C~41.17, shown in the upper right panel
of Fig.~\ref{4c41_fig}. The pixel size in this image ($\approx0.12\arcsec$)
is chosen to be $1/4$ of the native size of ACIS pixels. The image is smoothed
with a $0.37\arcsec\times0.37\arcsec$ top-hat kernel using the
\verb+aconvolve+ procedure. There appears to be an appreciable elongated structure
in the SW-NE direction, superimposed to the nuclear emission. We will proceed with
a quantitative analysis of this structure below. 

Overlaid on the \xray\ image are the contours of the 1.4~GHz radio emission, 
  generated from the NRAO VLA Archive Survey Images.\footnote{\url{https://archive.nrao.edu/nvas/}}
The $A$ and $B$ radio components defined in
  \citet{chambers+1990} and
  \citet{carilli+1994} are shown. The $A$ component can be seen as a compact radio knot at the position
  consistent with the edge of the elongated extended \xray\ emission.
  High-frequency radio images in \citet{carilli+1994}
reveal double hotspots within the $A$ component. The $B$ component is closely associated with the
core \xray\ emission. However, it is difficult to ascertain its physical nature based on the radio
map. While \citet{chambers+1990} argue that no flat-spectrum radio
core is detected, \citet{carilli+1994} claim the detection of radio core at 4.7~and
8.3~GHz, the position of which is labeled by the green cross in the upper-left panel of
Fig.~\ref{4c41_fig}. The
offset between the \xray\ and radio coordinates is $\approx0.6\arcsec$,
which is comparable to the typical 90\% uncertainty circle of the \chandra\
absolute position. We note, however, that the spectral index of the claimed radio
core is quite inverted ($\approx-0.7$ between 4.7 and 8.3 GHz), making the
identification of a self-absorbed, compact radio nucleus somewhat debatable.

PSF profiles were simulated using the \chandra\ Ray Tracer
\citep[ChaRT;][]{carter+2003}. Spectra of the central point source
(using a circular aperture of $2\arcsec$ radius) and the background
(using an annular region with the inner and outer radii of $10\arcsec$
and $15\arcsec$, respectively) were extracted using \verb+specextract+,
within {\sc sherpa} \citep{freeman+2001}. The saved spectra, together
with the corrected aspect solution file, were input to the ChaRT
server\footnote{\url{http://cxc.harvard.edu/ciao/PSFs/chart2/runchart.html}}. 
The output ray tracing file was then fed into the {\sc marx}
simulator\footnote{\url{http://space.mit.edu/cxc/marx/}} \citep{davis+2012}
to create the simulated PSF event file (shown in the lower right panel of
Fig.~\ref{4c41_fig}). To calculate the radial profile of the background-subtracted
source photon density, we adopted a series of concentric elliptical annular regions 
(see the contours in the 
lower panels of Fig.~\ref{4c41_fig}).\footnote{We acknowledge that without
  prior knowledge of the spatial profile of the extended emission, circular
  annuli should be the default choice over elliptical ones. We also calculated
  the radial photon density with circular annular regions and obtained consistent
  results.}, with semi-major
axes between $1\arcsec$ and $9\arcsec$, and 
steps of $1\arcsec$, and the semi-minor axes between $0.72\arcsec$
and $6.5\arcsec$, with steps of $0.72\arcsec$. The size parameters 
are chosen to ensure that the extended \xray\ emission can be covered by
a series of $\sim10$ annuli. Minor changes to these parameters will not
affect our results below on assessing the existence of extended emission.
The positioning angle of
the annuli is $325^\circ$, which is visually determined to be approximately along
the direction of the extended emission. 

A direct comparison of the radial profile of the source photon density profile 
for 4C~41.17 (solid line) against the simulated
PSF's (dashed line), shown in Fig.~\ref{4c41_ctsden_fig}, confirms the presence
of a statistically significant
extended component that goes well beyond $2\arcsec$ off the central point source
in 4C~41.17.
%Visually comparing the
%\chandra\ image with the radio maps of 4C~41.17 in \citet{chambers+1990} and
%\citet{carilli+1994} shows that
%the extended \xray\ emission is co-spatial with the extended radio emission. 

The measured photon density outside the central $2\arcsec$ declines following roughly an exponential law, and
becomes indistinguishable from the background level beyond $8\arcsec$.
To estimate the total \xray\ counts within the extended emission, we fit an exponential law
(corresponding to a linear relation if the photon density
is in log-scale) between $2\arcsec$ and
$8\arcsec$, and then extrapolated it into inside $2\arcsec$, yielding a total of $111.9^{+11.6}_{-10.6}$
counts over the full energy band. This was then subtracted from the total
\xray\ counts measured within a $8\arcsec$ radius to obtained the total point source counts, which
is $66.5^{+9.2}_{-8.1}$, where the quoted uncertainties are at the 
$1\sigma$ confidence level \citep{gehrels+1986}. 

Lastly, we analyzed the basic spectral properties of the extended emission and
the point source. We extracted the spectra from the (two) un-merged \chandra\ observations, 
and fit them jointly. Spectra for the extended emission were extracted from 
annular regions between $2\arcsec$ and $8\arcsec$, while the
spectra of the point source were extracted from circular regions within 
$2\arcsec$. The background spectra were extracted from annular regions 
between $10\arcsec$ and $15\arcsec$. The resulting spectra were binned to at least one count per bin (i.e., to eliminate zero
count channels), and 
analyzed with the {\sc xspec} package v.12.9.0 \citep{arnaud+1996} using 
$C$-statistic \citep{cash+1979}, which is best suited for low number count spectra \citep{nousek+1989}. 
For modelling purposes we adopted an absorbed power law  model (\verb+wabs+, \citealt{morrison+1983}), with hydrogen column density fixed at the Galactic value: $N_{\rm H} = 1.07\times10^{21}$~cm$^{-2}$ which is calculated using
the {\sc~colden} tool.\footnote{\url{http://cxc.harvard.edu/toolkit/colden.jsp} \citep{dickey+1990}. Including an extra component to account for intrinsic absorption and/or allowing 
  the hydrogen column density to vary did not yield any improvement in the fitting.}
The resulting photon index from the joint fitting is $\Gamma=1.6\pm0.4$, yielding a full-band,  observer-frame flux of $F_{\rm X} = 6.56^{+0.68}_{-0.62}\times10^{-15}$~erg~cm$^{-2}$~s$^{-1}$ for the extended emission. The fitting statistic $C/n=89.4/93$, where $C$ is the $C$-statistic and $n$ is the number of bins. 
The spectra of the point source were fit using the same model. The resulting
\xray\ photon index, $\Gamma=0.8\pm0.3$, is much harder than that of the extended emission, possibly indicating strong intrinsic
absorption, or a different emission mechanism. The full-band flux is
$F_{\rm X} = 6.48^{+0.89}_{-0.79}\times10^{-15}$~erg~cm$^{-2}$~s$^{-1}$. For the point source, $C/n=91.9/102$. 
%Flux
%densities in the four sub-bands are also listed in Table~\ref{cts_table}. 

\subsection{4C~03.24}\label{data:4c03}

\begin{figure*}
    \centering
    \includegraphics[width=6.5in]{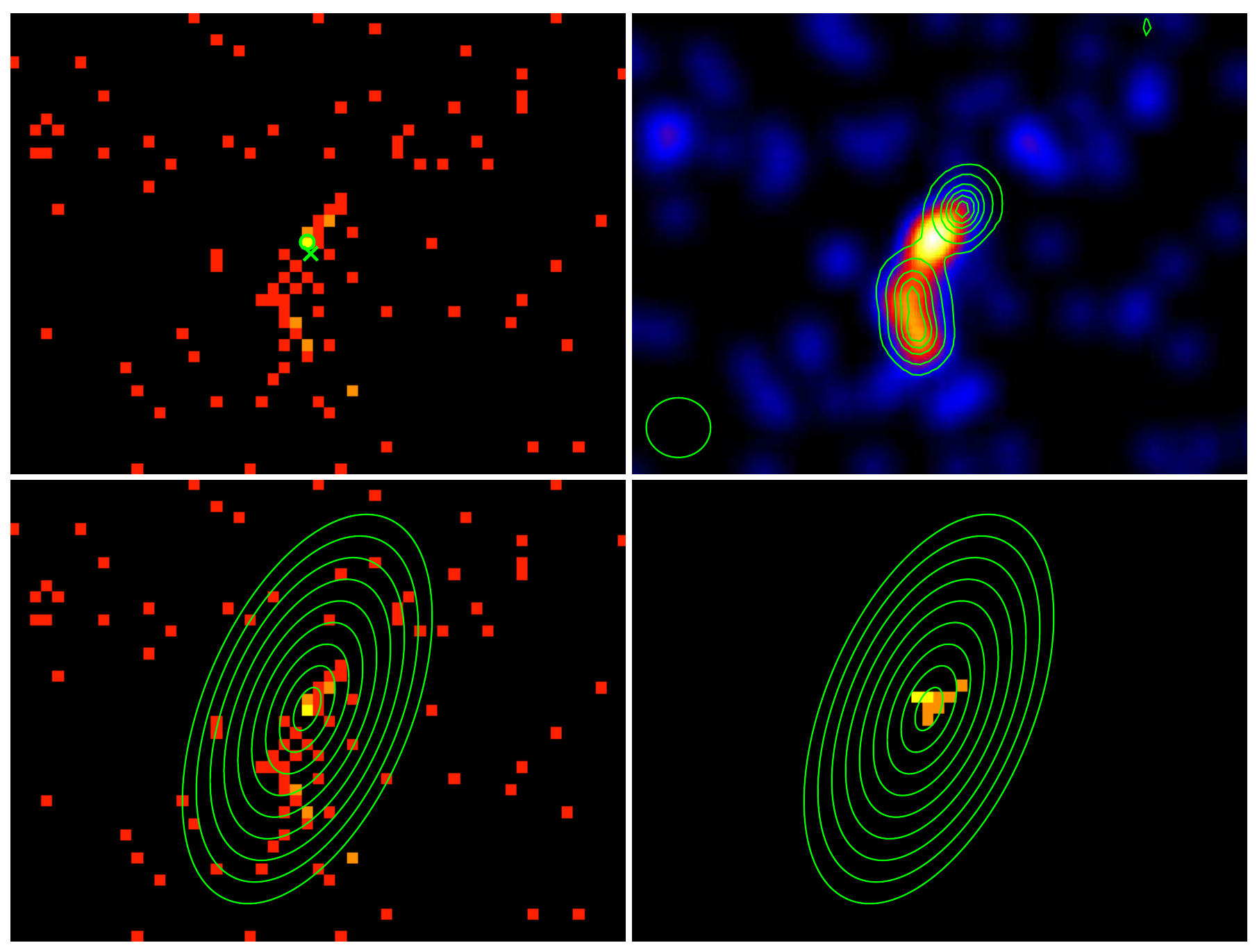}
    \caption{
      The \chandra\ ACIS images of 4C~03.24. The left two
        panels are the original full-band (0.5--8.0~keV) images. The upper right
        panel is the smoothed subpixel images, with the overlaid 1.4~GHz radio
        flux contours. The contour levels are 2, 10, 25, 40, 65, and 90 mJy/beam.
        The beam size is shown as
        the ellipse in the bottom left corner of this panel, with semi-major axis of
        $1.4^{\prime\prime}$ and semi-minor axis of $1.3^{\prime\prime}$.
        The lower right panel is the ChaRT
        simulated PSF for the \chandra\ observation. In the upper left panel, the
        open circle and the cross symbol label the \xray\ position and the radio position
        of the point source, respectively. The elliptical annular regions are those
        used in calculating the profile of the \xray\ count number density for the
        extended \xray\ emission.} 
    \label{4c03_fig}
\end{figure*}

A similar procedure to what described above for 4C~41.17 was adopted to analyze the ACIS data of 4C~03.24, consisting of one observation with (good) exposure time of 90.6~ks. The full-band, raw \xray\ image and the smoothed subpixel ($0.12\arcsec$) image are shown in the top panels of Fig.~\ref{4c03_fig}. A point \xray\ source is clearly detected. Extended \xray\ lobes appear to be along the NW-S direction.

The 1.4~GHz radio contours are also shown in the upper right panel, overlaid on the smoothed \xray\ image. The extended radio emission along the S-direction is aligned with the extended \xray\ emission, while the radio knot along the NW-direction near the edge of the \xray\ lobe can possibly be considered as a radio hotspot. \citet{vanojik+1996} provided 1.5~GHz, 4.7~GHz, and 8.3~GHz radio maps (see their Figs.~3 and 4). A radio core is detected at 4.7~GHz and 8.3~GHz, albeit with a steep spectral index, too ($\approx-1.2$). This radio core is $\approx0.5\arcsec$ away from the point \xray\ source.

To assess the statistical significance of the extended \xray\ emission,
we defined a series of concentric elliptical annuli with semi-major axes between
$1\arcsec$ and $9\arcsec$ and steps of $1\arcsec$, and semi-minor axes between $0.5\arcsec$ and $4.5\arcsec$ with steps
of $0.5\arcsec$\ (see the lower panels of Fig.~\ref{4c03_fig}). The positioning angle is $67^\circ$.
The PSF was again simulated with ChaRT and {\sc marx}, following the same procedure as outlined above (lower right panel of Fig.~\ref{4c03_fig}). The radial photon density profiles
of 4C~03.24 is compared to the simulated PSF's in Fig.~\ref{4c03_ctsden_fig}, showing evidence for extended \xray\ emission beyond $1\arcsec$. The total, full-band counts for the extended emission and the point-source are $34.1^{+6.9}_{-5.8}$ and $4.6^{+3.3}_{-2.1}$, respectively.

As for 4C~41.17, we extracted and analyzed the extended emission
and point source spectra; the hydrogen column density was fixed at the 
Galactic value: $N_{\rm H} = 1.92\times10^{20}$~cm$^{-2}$. For the extended
emission, the best-fitting photon index is $\Gamma=1.4\pm0.6$ ($C/n=17.9/25$). The unabsorbed full-band 
flux is $F_{\rm X} = 4.69^{+0.93}_{-1.27}\times10^{-15}$~erg~cm$^{-2}$~s$^{-1}$. 
For the point source, $\Gamma=1.2^{+0.8}_{-0.4}$ ($C/n=5.1/13$). The unabsorbed
full-band flux is $F_{\rm X} = 2.41^{+1.02}_{-0.87}\times10^{-15}$~erg~cm$^{-2}$~s$^{-1}$.
%The large error bars of the photon indices and flux values are due to the
%very limited \xray\ counts.
For both 4C~41.17 and 4C~03.24, it is
not feasible to decompose the diffuse emission component from the core
source inside the $2^{\prime\prime}$-radius circular region due to insufficient counts. 

\subsection{4C~19.71, 4C~60.07 and \tnj }\label{data:other}

\begin{figure*}
    \centering
    \includegraphics[width=5.5in]{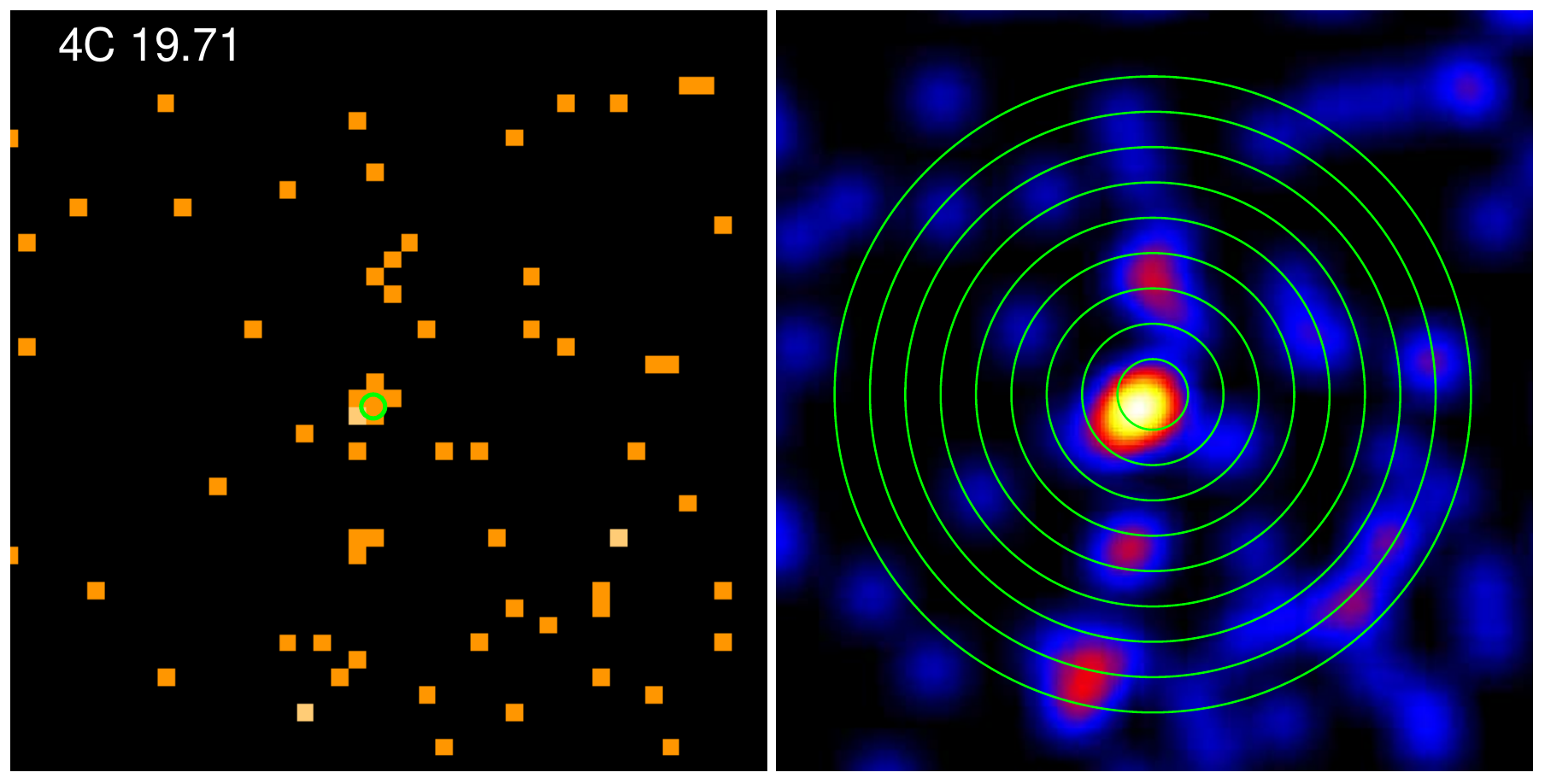}\\
    \includegraphics[width=5.5in]{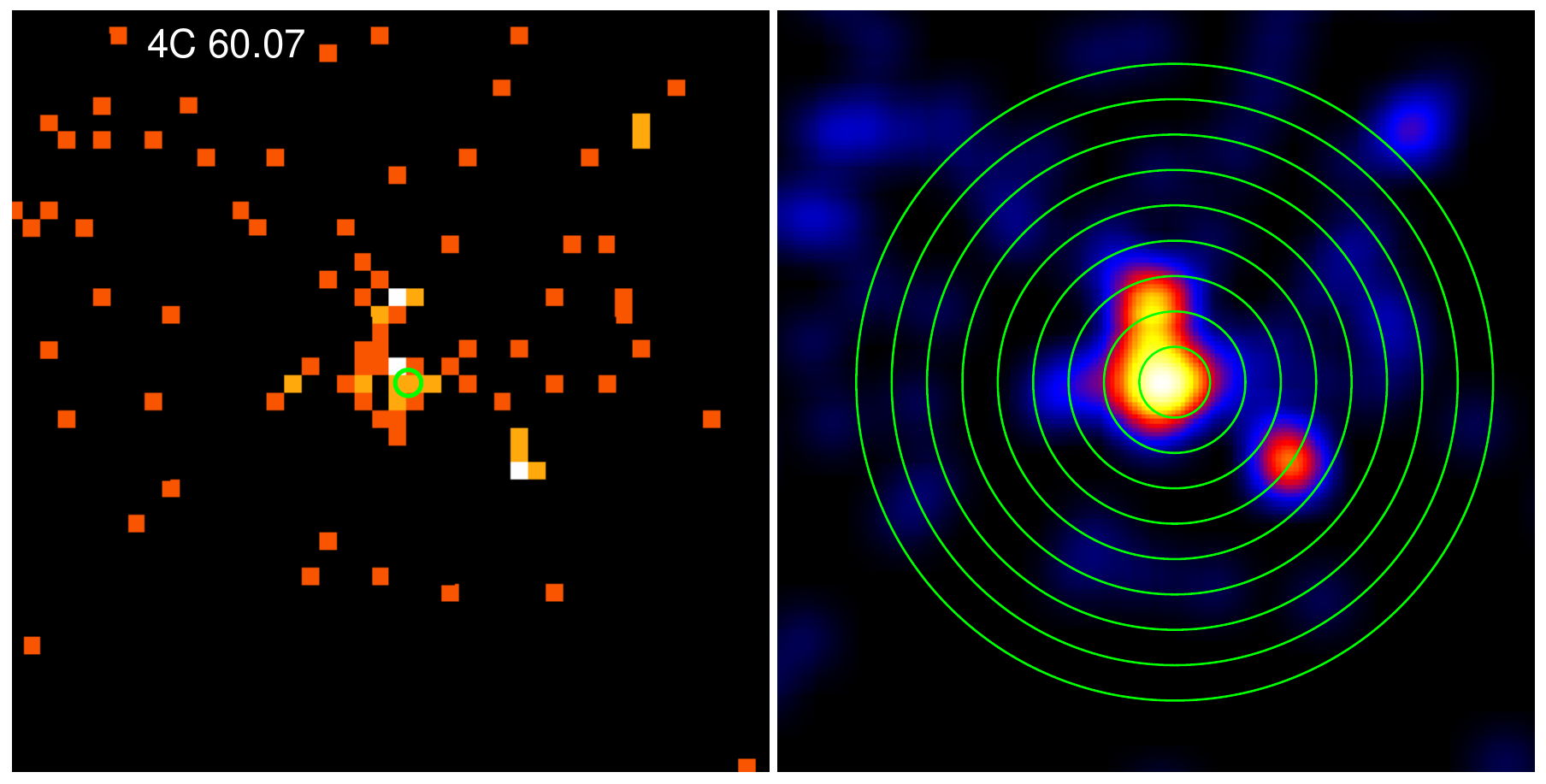}\\
    \includegraphics[width=5.5in]{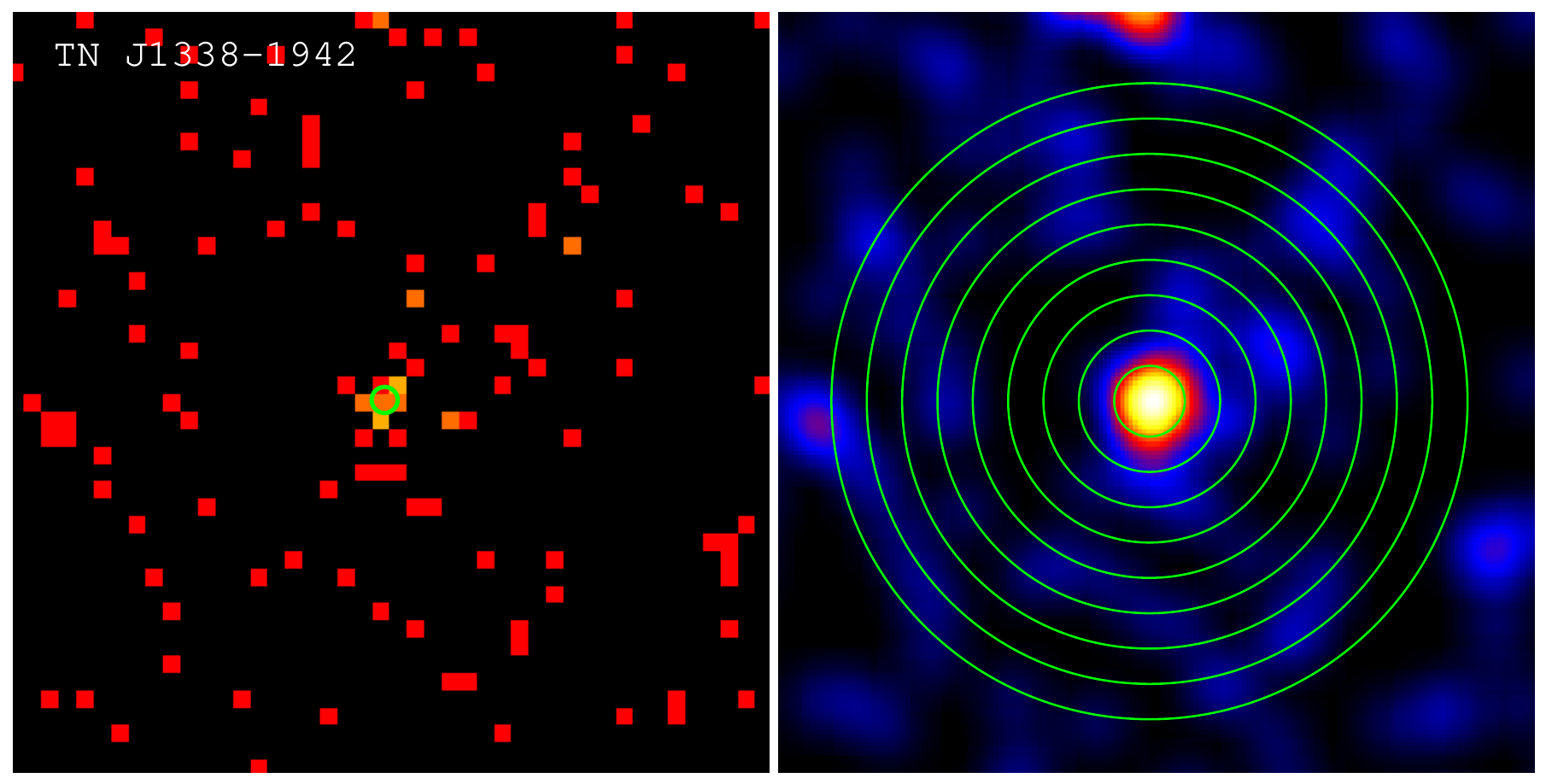}\\
    \caption{
      \chandra\ ACIS images of  4C~19.71 (top), 4C~60.07 (middle), and \tnj\ (bottom). For
      each object, the raw,  full-band (0.5--8.0~keV) image is shown on the left, while the smoothed subpixel image is shown on the right. The open circles in the left panels label the position of the \xray\ point sources.
      The green annular regions in each right panel are those
        used in calculating the profile of the \xray\ count number density for the
        extended \xray\ emission. None of thee objects shows evidence for a statistically significant extension in the \xray\ emission. \label{other_fig}}
\end{figure*}

\begin{figure*}
 \centering
    \includegraphics[width=6.5in]{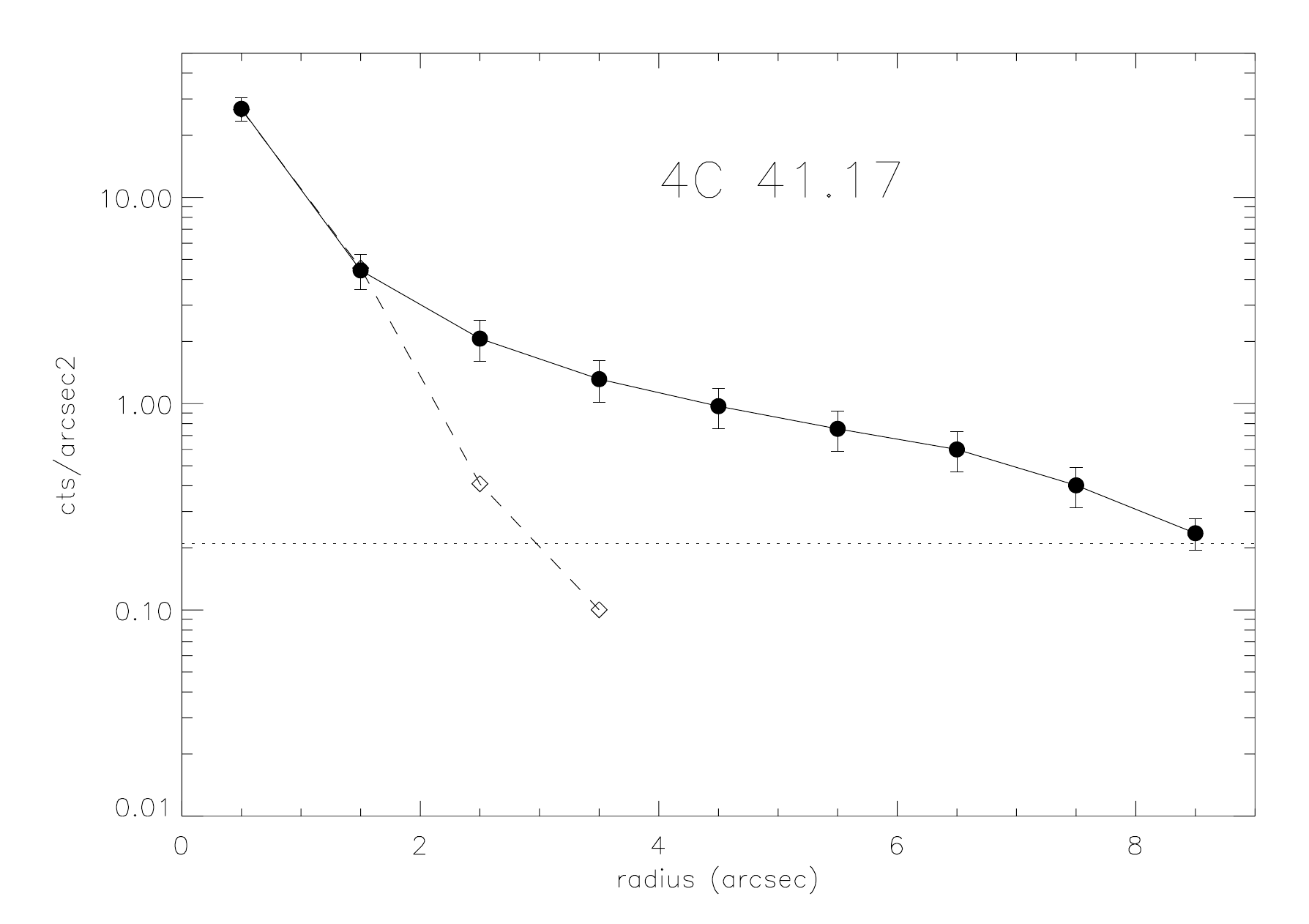}
    \caption{The photon density profile of 4C 41.17 (solid line) is shown vs. 
        the simulated PSF's (dashed line). The source's profile is calculated along the major axes of the
        elliptical regions shown in Fig.~\ref{4c41_fig}. The dotted horizontal line indicates the
        background value. \label{4c41_ctsden_fig}}
\end{figure*}

\begin{figure*}
    \centering
    \includegraphics[width=6.5in]{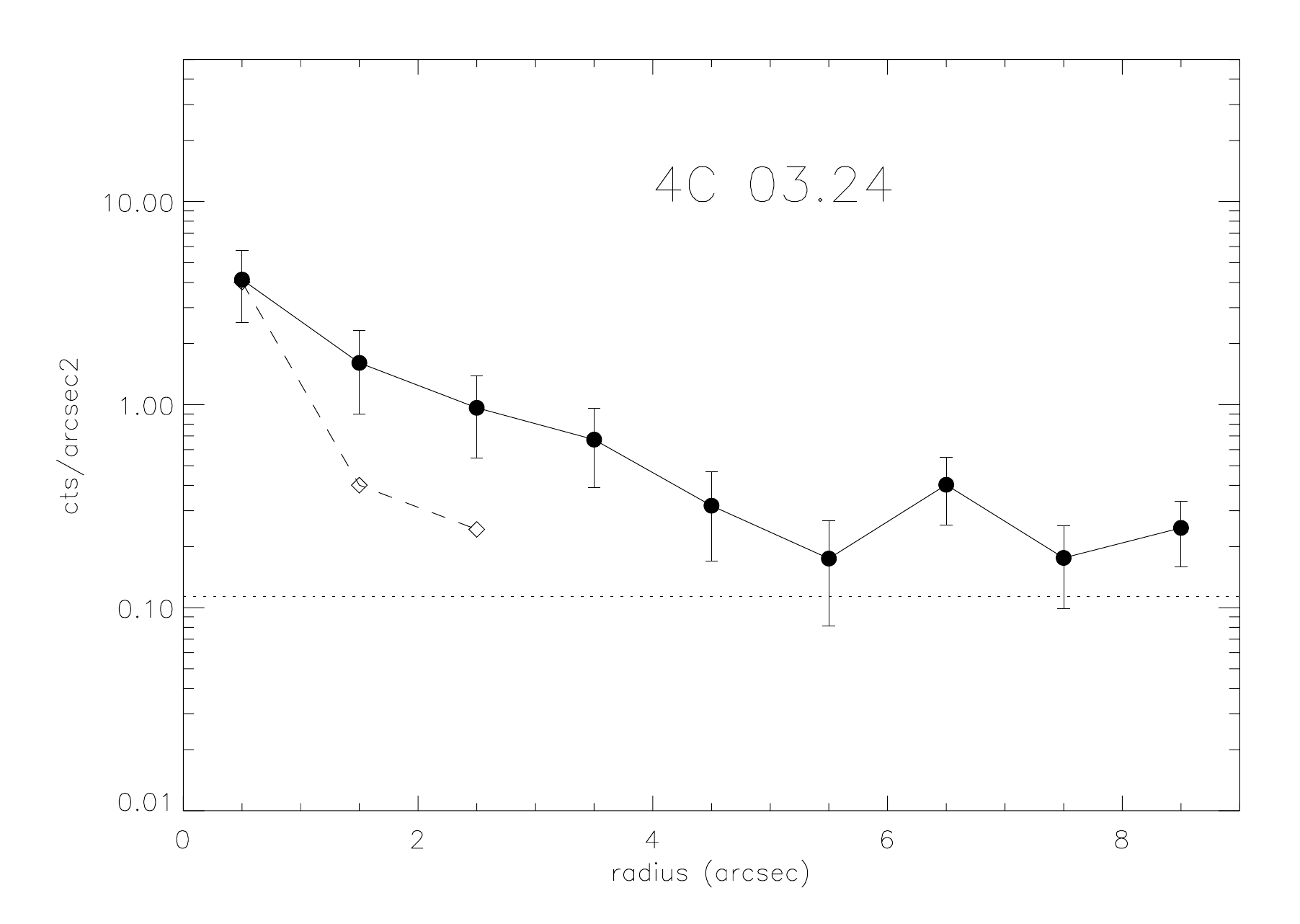}
    \caption{Photon density profile of 4C 03.24 (solid line) vs. 
        the simulated PSF's (dashed line). The source's profile is calculated along the major axes of the
        elliptical regions shown in Fig.~\ref{4c03_fig}. The dotted horizontal line indicates the
        background value.\label{4c03_ctsden_fig}}
\end{figure*}

The full-band and sub-pixel images of the remaining three high-redshift
radio galaxies with deep \chandra\ data are presented in Fig.~\ref{other_fig}: as discussed next, contrary to what reported in the literature listed in Table~\ref{log_table}, {\it we could not confirm the presence of statistically significant extended \xray\ emission associated with the radio lobes in these objects}.  

\begin{figure*}
    \includegraphics[width=4.0in]{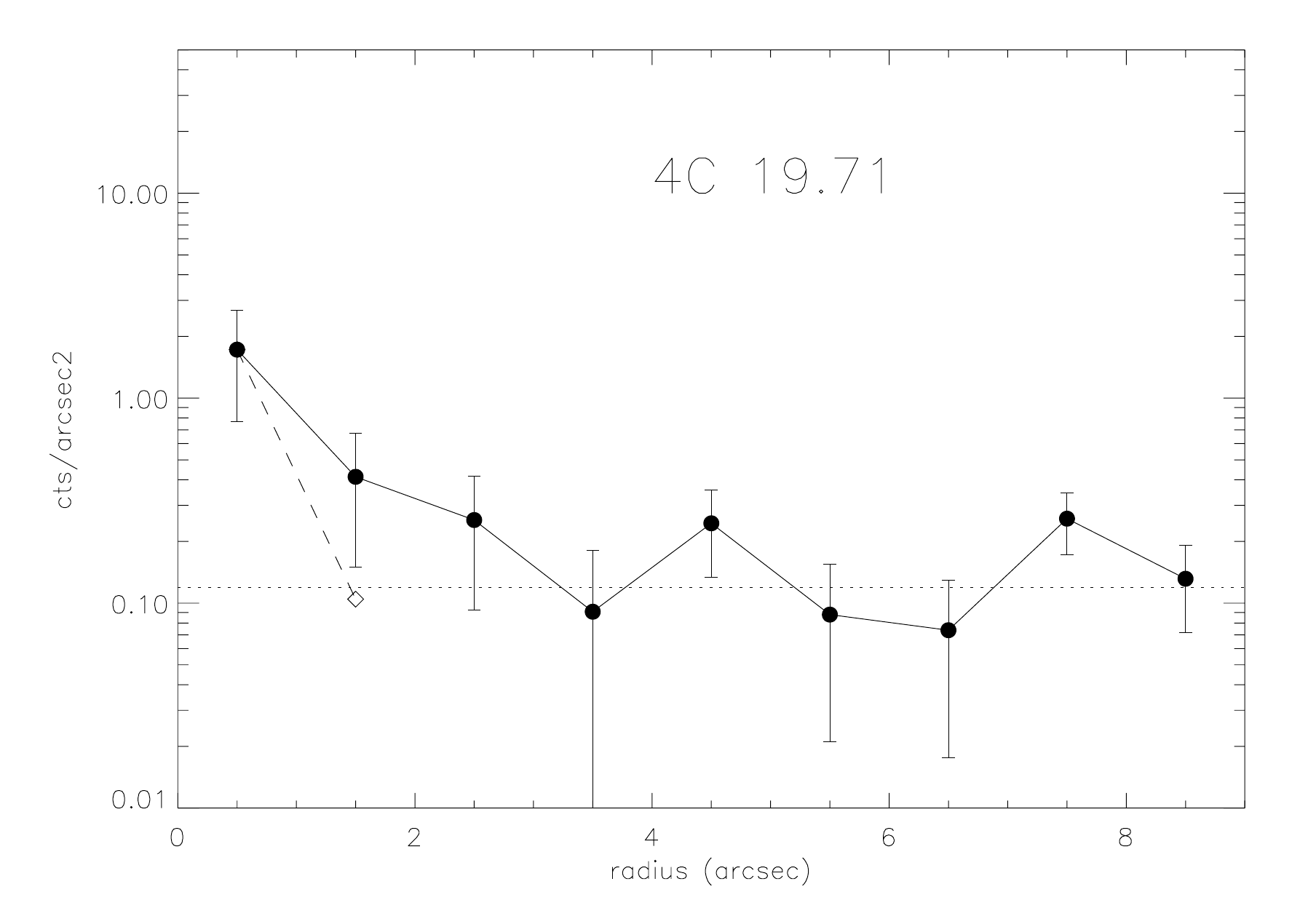}\\%\vspace{-6cm}
    \includegraphics[width=4.0in]{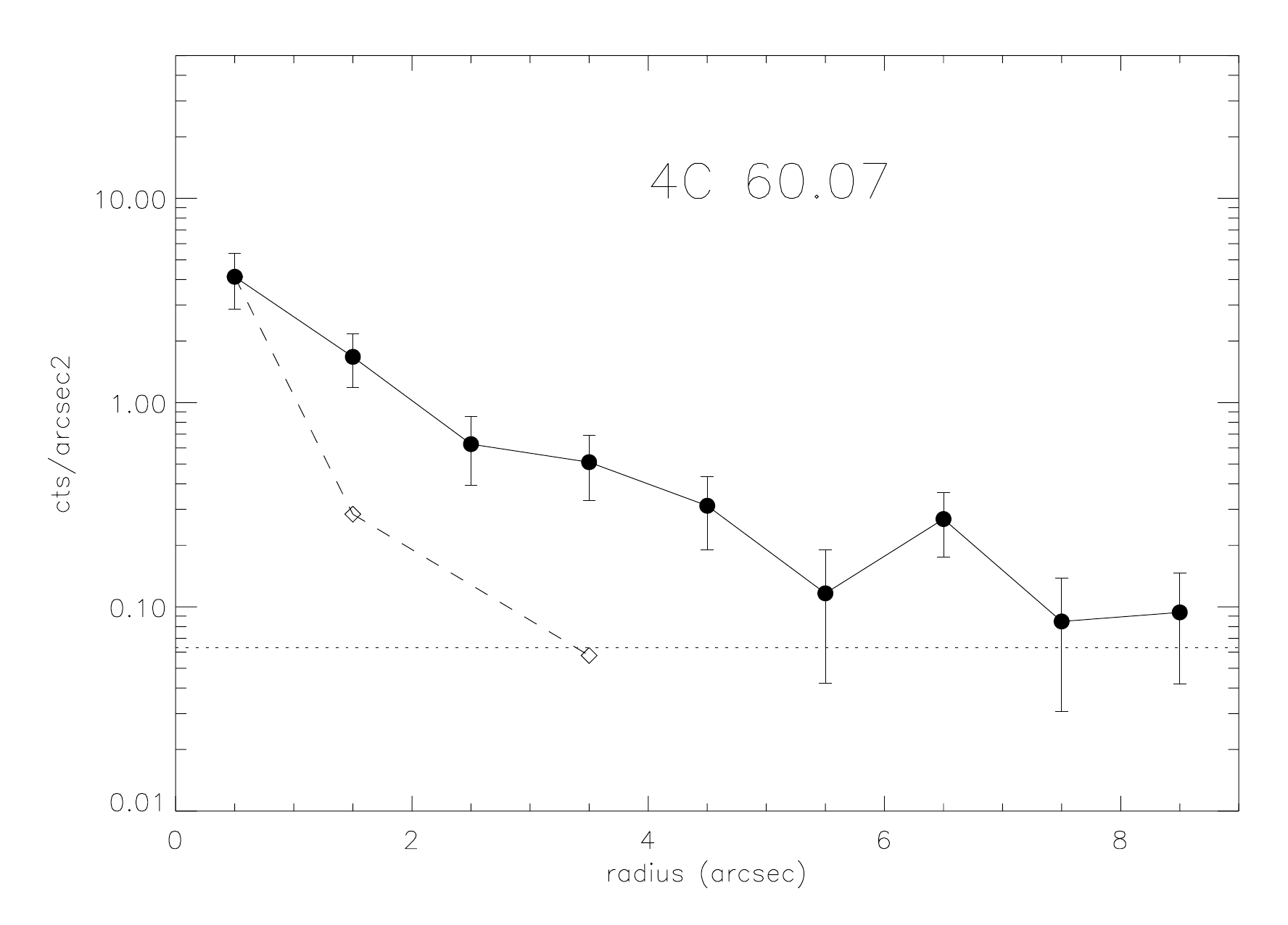}\\%\vspace{-6cm}
    \includegraphics[width=4.0in]{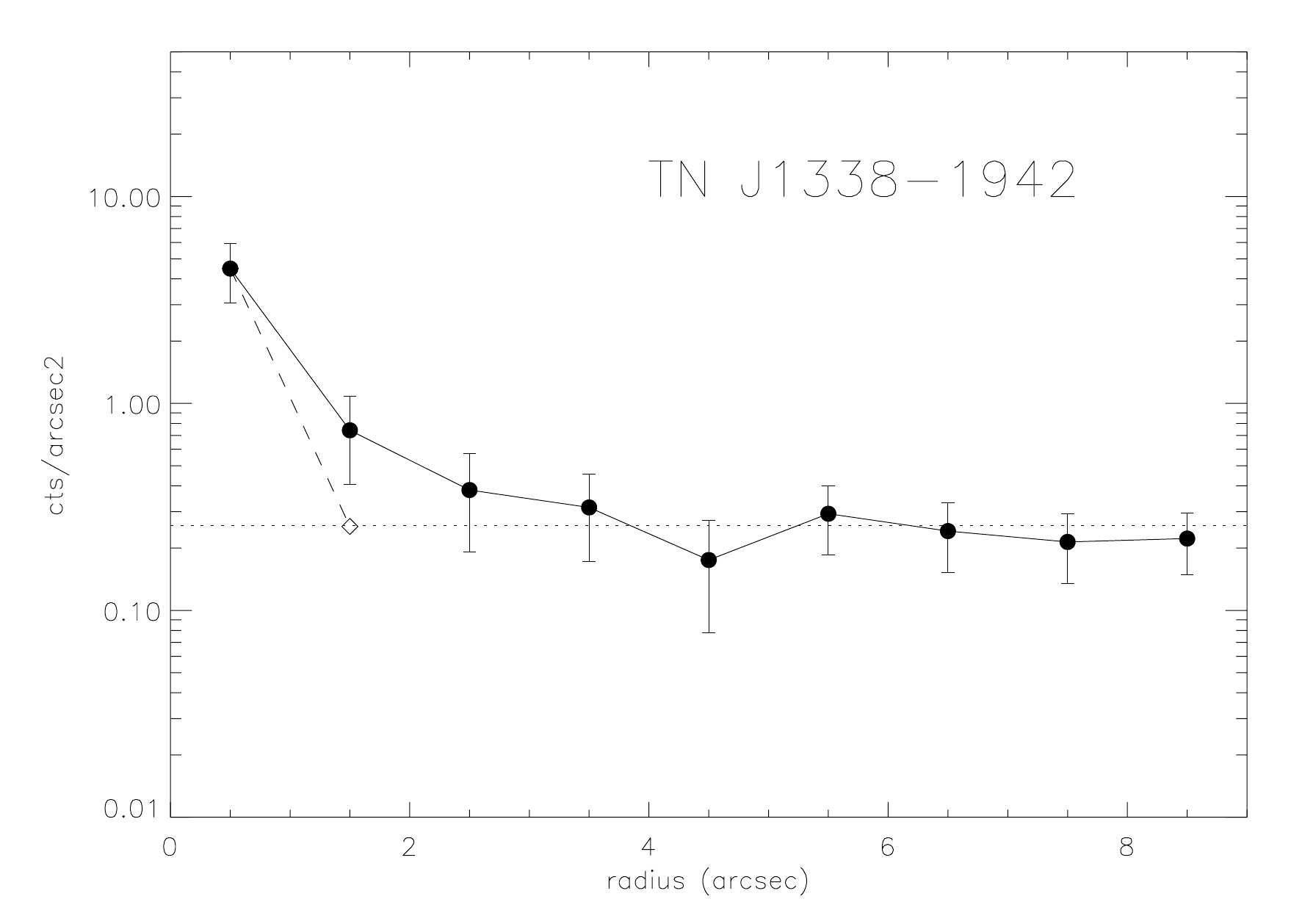}\\%\vspace{-4cm}
    \caption{The radial photon density profiles for 4C~19.71, 4C~60.07,
        and \tnj\ (solid lines), and the corresponding simulated PSF (dashed lines).
        The horizontal dotted line in each panel represents the photon density of the
        background. The profiles of the radio galaxies are calculated along the radii of the
        annular regions shown in the right panels of Fig.~\ref{other_fig}}
    \label{other_ctsden_fig}
\end{figure*}

A bright, point-like \xray\ source was detected at a position consistent with the optical source position in each target. In line with the previous two cases, we derived the sources' radial photon density profiles and compared them to the simulated PSFs' profiles (see Fig.~\ref{other_ctsden_fig}).
For 4C~19.71, the detected \xray\ source's radial profile 
decreases so rapidly that it is consistent with background beyond $2\arcsec$, similarly to
that of the simulated PSF's (top panel of Fig.~\ref{other_ctsden_fig}). The \xray\ image might suggest the presence of 
possible, off nuclear \xray\ sources located $4\arcsec$ and $5\arcsec$ away from the central point
source in the N and S direction, respectively (see the top right panel of Fig.~\ref{other_fig}).
The detection of co-spatial 8.3~GHz radio lobes at these positions \citep{smail+2012} admittedly 
increases the probability that such extended \xray\ emission is indeed present. 
However, from a statistical perspective, none of the visual enhancements represents a significant detection above the background. 

%only a minimal enhancement of the associated
%with the center point source. However, the extended emission has very limited
%counts (barely increasing the photon density above the background level; see upper
%panel of Fig.~\ref{other_ctsden_fig}), which would produce large uncertainties to
%its flux density.

For 4C~60.07, the radial photon density exceeds the background level within $\sim 5\arcsec$
from the central \xray\ source position (middle panel of Fig.~\ref{other_ctsden_fig}). This excess emission appears to be 
mainly composed by two off-nuclear sources, respectively $\approx 3\arcsec$ away in the N direction and 
$\approx5\arcsec$ away in the SW direction (see the middle right panel of
Fig.~\ref{other_fig}). However, these sources are located in a direction which is nearly {\it perpendicular} to that of the galaxy 
radio lobes (see Fig.~1 of \citealt{smail+2009}). At the same time, the measured \xray\ counts within the region corresponding to the location of the radio lobes is consistent with the background level. 

The radial photon density profile of \tnj\ is entirely consistent with the PSF's profile 
(bottom panel of Fig.~\ref{other_ctsden_fig}). It reaches the background level beyond
$2\arcsec$. 

In summary, we find no compelling, statistically significant evidence for extended \xray\
emission that is co-spatial with the known radio lobes in any of the three sources considered
here (albeit marginal evidence is found for 4C~19.71). For the SED modelling purposes,
the \xray\ upper limits cannot provide meaningful constraints on their SEDs. We shall then
focus on 4C~41.17 and 4C~03.24, for which we demonstrated the presence of significant,
extended \xray\ emission spatially associated with the known radio lobes. It is worth
noting that the non-detection of extended \xray\ emission does not necessarily rule out
the existence of \xray\ jets and/or lobes (e.g., see discussion in \citealt{fabian+2014}).

One caveat is that the above methodology of using annular regions to verify the
existence of extended \xray\ emission has its limits. It is most effective at searching
for diffuse \xray\ emission superimposed upon a center point source. However, in case
of two or more well separated point sources, e.g., the core and the jet knots, it is
difficult to reach definitive conclusion without visually inspecting the \xray\
image. None of the five objects examined here suffer from this potential issue except
4C~60.07, for which, as discussed above, the two point \xray\ sources in addition to
the center core are not likely associated with the radio galaxy. 

% Table~2: X-ray Photometry
\begin{table*}
\centering
\begin{tabular}{ccccc}
%\tablecaption{X-ray Photometry for the High-$z$ Radio Galaxies\label{cts_table}}
  %\tablewidth{0pt}
  \hline
  \hline
 {Name} &  {Band} & {Energy (keV)/} & \multicolumn{2}{c}{X-ray flux/flux density}  \\
  \cline{4-5}
        {} & {} & {Frequency (Hz)} & {Extended} & {Point Source} \\
        \hline
%\startdata
4C~41.17 & Full-Band & 0.5--8.0 & $6.56^{+0.68}_{-0.62}$ & $6.48^{+0.89}_{-0.79}$ \\
%\cline{2-5}
& B1 & $1.71\times10^{17}$ & $7.0\pm1.5$ & $0.7\pm0.6$ \\
& B2 & $3.42\times10^{17}$ & $6.0\pm0.9$ & $3.2\pm0.7$ \\
& B3 & $6.84\times10^{17}$ & $3.7\pm0.9$ & $4.7\pm1.0$ \\
& B4 & $1.37\times10^{18}$ & $5.5\pm1.4$ & $3.5\pm1.2$ \\
\hline
4C~03.24 & Full-Band & 0.5--8.0 & $4.74^{+0.96}_{-0.81}$ & $0.70^{+0.50}_{-0.32}$ \\
%\cline{2-5}
& Soft-Band & $2.42\times10^{17}$ & $5.3\pm1.1$ & $0.5\pm0.4$ \\
& Hard-Band & $9.67\times10^{17}$ & $1.6\pm0.6$ & $0.4\pm0.4$ \\
%\enddata
\hline
\hline
\end{tabular}
\caption{X-ray Photometry for the High-$z$ Radio Galaxies.
  The numbers for ``Full-Band'' are the \xray\ flux, in unit of $10^{-15}$
  erg~cm$^{-2}$~s$^{-1}$. The numbers for the sub-bands are the \xray\ flux density at a given
  frequency, in unit of $10^{-33}$ erg~cm$^{-2}$~s$^{-1}$~Hz$^{-1}$.
}
\label{cts_table}
\end{table*}

% -----------------------------------------------------------------------------
% SEDs
% -----------------------------------------------------------------------------

%\section{Spectral Energy Distributions}\label{sed}

\section{Observed SEDs}\label{makesed}

%In order to
%construct the spectral energy distribution (SED) of 4C~41.17 (see \S\ref{modelsed}),
%we broke the full band into four sub-bands (B1: 0.5--1.0~keV; B2: 1.0--2.0~keV;
%B3: 2.0--4.0~keV; B4: 4.0--8.0~keV) and calculated the flux density at the
%geometric mean frequency of each sub-band (see Table~\ref{cts_table}).

%Both the extended and the point source emission from 4C~03.24 have
%significantly fewer counts than those of 4C~41.17. Therefore,
%we are only able to break the full band into two sub-bands, the soft band
%(0.5--2.0~keV) and hard band (2.0--8.0~keV). The flux density at the center of each
%sub-band is also listed in Table~\ref{cts_table}. 

\begin{figure*}
    \centering
    \includegraphics[width=6.5in]{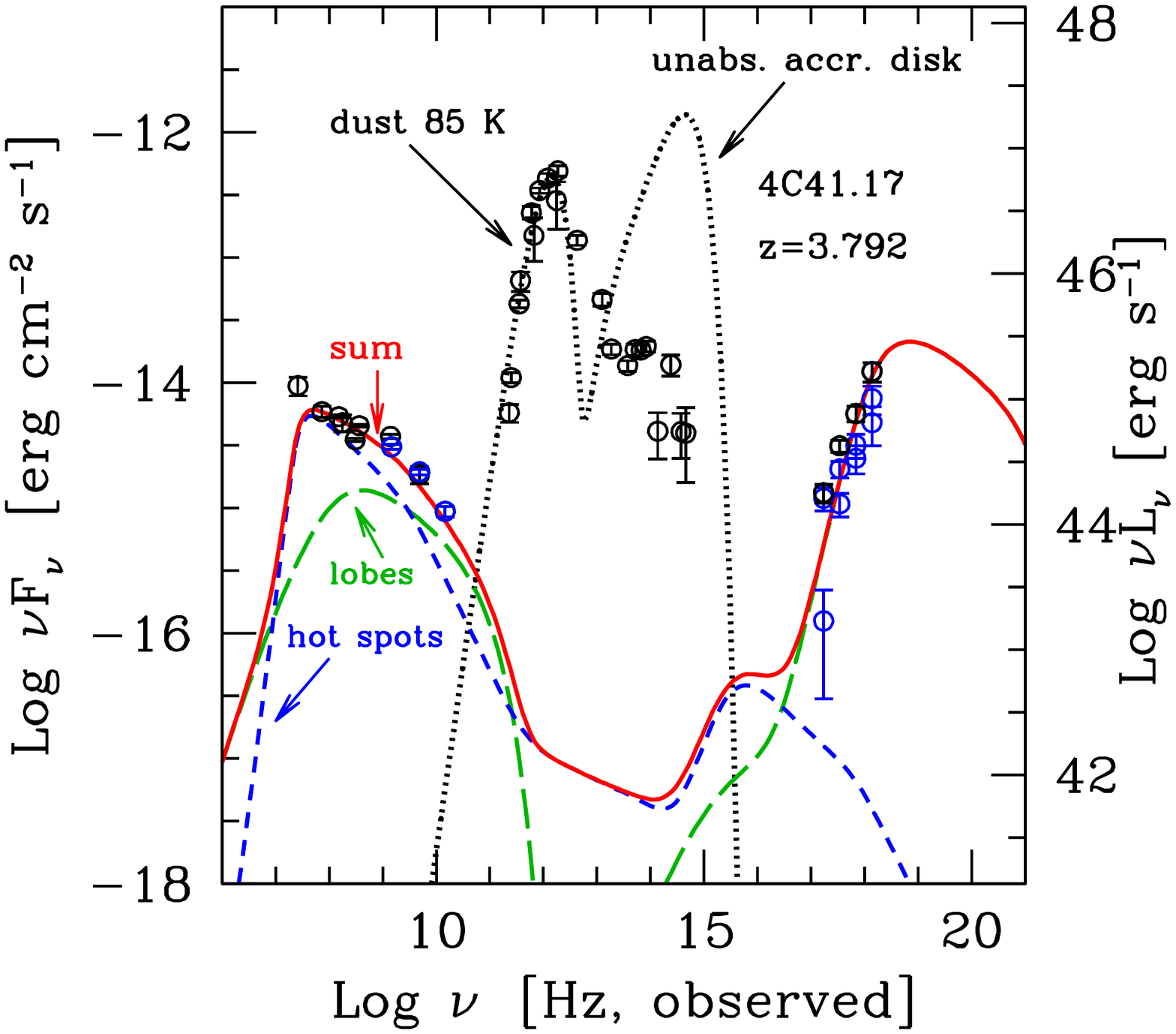}
    \caption{The SED of 4C~41.17 and the applied models. The open circles
      (and error bars) are the observed SED data points. The black dotted line
      corresponds to contribution of an accretion disk and absorbing dust,
      re-emitting in the IR as a blackbody of temperature $T_{\rm IR}$=85~K.
      This material can correspond to an absorbing torus close to the accretion
      disk and/or some extra absorbing material in the host galaxy at large
      distances from the disk. We show the {\it unabsorbed} disk emission,
      that corresponds to a luminosity $L_{\rm d}=2.6\times10^{47}$ erg s$^{-1}$.
      The inverse Compton flux from the jet is below the scale of the figure.
      The short-dashed blue line correspond to the emission from the hotspots
      (with parameters listed in Table~\ref{paralobe_table}). It contributes
      mainly to the low frequency radio emission. The long-dashed green line
      is the flux produced by the lobes (parameters in
      Table~\ref{paralobe_table}) contributing to the high frequency radio
      emission and especially to the \xray\ flux. The solid red line is the
      sum of the hotspot and lobe fluxes.}
    \label{4c41_sed_fig}
\end{figure*}

\begin{figure*}
    \centering
    \includegraphics[width=6.5in]{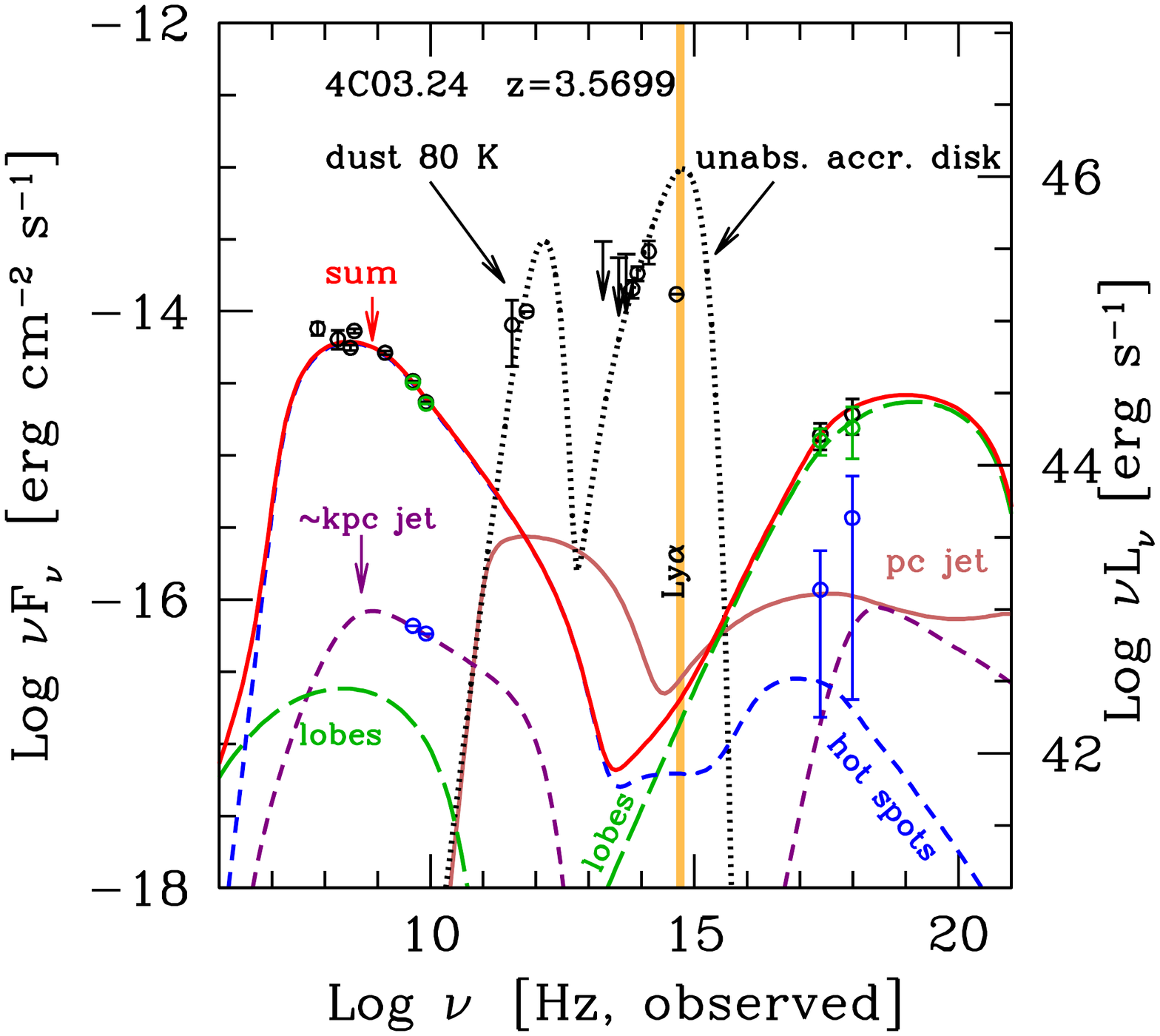}
    \caption{The SED of 4C~03.24 and the applied model. The open circles
      (and error bars) are the observed SED data points. The black dotted line
      corresponds to contribution of an accretion disk and absorbing dust,
      re-emitting in the IR as a blackbody of temperature $T_{\rm IR}$=80~K.
      We show the {\it unabsorbed} disk emission, that corresponds to a
      luminosity $L_{\rm d}=1.5\times10^{47}$ erg s$^{-1}$. The vertical orange
      line labels the observed-frame frequency of Ly$\alpha$ emission.
      The solid brown
      line is the emission produced by the inner (pc) relativistic jet, with
      parameters listed in Table~\ref{para_table}. 
The synchrotron component of this emission is hidden below the dust and the disk
emission, while the inverse Compton flux could contribute to the X--ray flux
of the core. 
The violet dashed line corresponds to the emission from the jet, but at
much larger ($\sim$kpc) scale. It is responsible for the relatively steep radio emission
observed at the $\sim$ arcsec scale.
The short-dashed blue
      line corresponds to the emission from the hotspots (with parameters
      listed in Table~\ref{paralobe_table}). It contributes mainly to the
      radio emission. The long-dashed green line is the flux produced by
      the lobes (parameters in Table~\ref{paralobe_table}) contributing
      to the \xray\ flux. The solid red line is the sum of the
      hotspot and lobe fluxes.}
    \label{4c03_sed_fig}
\end{figure*}

The SEDs of 4C~41.17 and 4C~03.24 are shown in Fig.~\ref{4c41_sed_fig} and
Fig.~\ref{4c03_sed_fig}, respectively, where the total emission corresponds to the black open circles, whereas, whenever a decomposition was possible, the point source and extended emission are marked by blue and green circles, respectively.
All the circle symbols represent observed data, while the modelling of the SEDs are detailed in the next section. 
The nature of the point radio sources is discussed in detail below. 
For the \xray\ data points, the full band was broken into four sub-bands for 4C~41.17 (B1: 0.5--1.0~keV; B2: 1.0--2.0~keV;
B3: 2.0--4.0~keV; B4: 4.0--8.0~keV), where the flux densities are expressed at the
geometric mean frequency of each sub-band (see Table~\ref{cts_table}). Since both the extended and the point source component in 4C~03.24 have significantly fewer counts than those in 4C~41.17, for this system we broke the full band into the two standard sub-bands: the soft band
(0.5--2.0~keV) and hard band (2.0--8.0~keV; again see Table~\ref{cts_table}). 

Flux densities in the radio, sub-millimeter, infrared and optical bands
were retrieved from the NASA/IPAC Extragalactic Database
(NED)\footnote{\url{http://ned.ipac.caltech.edu}}. For most of these SED
data points, only the total flux is available. Owing to the limited angular resolution of the corresponding observations, we typically only tabulate the total flux densities at each available frequency.  However, high angular resolution radio maps are available for both 4C~41.17 and 4C~03.24, enabling us to attempt a core vs. lobe decomposition at GHz frequencies. This is briefly discussed next. 

As discussed in \S\ref{data:4c41}, 4C~41.17 has several components of extended
radio emission,
while the existence of a compact radio core is debatable. Regardless, the flux density of
the alleged radio core is substantially lower (by a factor of $>40$) than
that of the extended radio emission, thus having negligible effects on the estimated
total flux density. Therefore, for SED modelling purposes, we choose to adopt the radio
flux densities at 1.5~GHz, 4.9~GHz, and 14.7~GHz given by \citet{chambers+1990}. 
%The radio core flux densities given by \citet{carilli+1994} are also shown in
%Fig.~\ref{4c41_sed_fig} (red symbols) for reference.

For 4C~03.24, radio flux densities at 4.7~GHz, and 8.3~GHz are retrieved
from Table~2 of \citet{vanojik+1996}, in which the $N$ component is considered as the
``radio core'', while other components combined as the extended radio emission. 
%The fluxes of the radio lobes and core
%are represented by the blue and red symbols, respectively. 

%------------------------------------------------
\begin{table*} 
\centering
\begin{tabular}{l l l l l l l l l l l l l l l l l l}
\hline
\hline
Name &$z$ &$M$ &$L_{\rm d}$ &$R_{\rm diss}$ &$R_{\rm BLR}$ &$P^\prime_{\rm e, jet, 45}$  &$B$ &$\Gamma$ &$\theta_{\rm V}$  
  &$\gamma_{\rm b}$ &$\gamma_{\rm max}$ &$s_1$ &$s_2$    &$\log P_{\rm jet}$ \\ % &$\log P_{\rm r}$
~[1] &[2] &[3] &[4] &[5] &[6] &[7] &[8] &[9] &[10] &[11] &[12] &[13] &[14]  &[15]  \\
\hline   
4C 41.17        &3.792 &5e9  &260   &2e3 &1.6e3 &0.01 &1.6   &12 &45 &100 &4e3 &1    &3.4  &47.2  \\ % &45.5  
4C 03.24 ``pc"  &3.57  &6e8  &11.7  &108 &342   &0.06 &5.3   &10 &17 &100 &3e3 &0    &2.0  &46.9   \\  
4C 03.24 ``kpc" &3.57  &6e8  &11.7  &1.8e6 &342   &0.1  &2e--4 &3  &17  &600 &4e4 &--1 &3.45 &45.7   \\  
\hline
\hline 
\end{tabular}
\vskip 0.4 true cm
\caption{
Adopted parameters for the jet models shown in Fig. \ref{4c41_sed_fig} and Fig. \ref{4c03_sed_fig}.
Col. [1]: name; 
Col. [2]: redshift;
Col. [3]: black hole mass in solar masses;
Col. [4]: disk luminosity in units of $10^{45}$ erg s$^{-1}$;
Col. [5]: distance of the dissipation region from the black hole, in units of $10^{15}$ cm;
Col. [6]: size of the BLR, in units of $10^{15}$ cm;
Col. [7]: power injected in the jet in relativistic electrons, calculated in the comoving 
frame, in units of $10^{45}$ erg s$^{-1}$;
Col. [8]: magnetic field in G;  
Col. [9]: bulk Lorentz factor;
Col. [10]: viewing angle in degrees;
Col. [11] and Col. [12]: break and maximum Lorenz factor of the injected electron distribution;
Col. [13] and Col. [14]: slopes of the injected electron distribution; 
Col. [15]: logarithm of the total kinetic plus magnetic jet power, in erg s$^{-1}$.
% Col. [16]: logarithm of jet power in the form of radiation, in erg s$^{-1}$;
The values of the powers and the energetics refer to {\it one} jet.
}
\label{para_table}
\end{table*}
% --------------------------------------

%------------------------------------------------
\begin{table*} 
\centering
\begin{tabular}{l l l l l l l l l l l l l l l l l}
\hline
\hline
Name &Comp. &$R$ &$\log P_{\rm e}$ &$B$  &$\gamma_{\rm b}$ &$\gamma_{\rm max}$ &$s_1$ &$s_2$ 
&$\log E_{\rm e}$  &$\log E_{\rm B}$   \\ 
~[1] &[2] &[3] &[4] &[5] &[6] &[7] &[8]&[9] &[10] &[11]\\
\hline   
4C 41.17 &HS      &3.9  &46.7 &9.0e--4 &100  &1e5  &--1 &3.8 &58.0 &59.4   \\  
         &lobe    &48.7 &46.1 &1.7e--5 &4e3  &1e5  &--1 &2.8 &59.3 &59.2   \\ 
%4C 03.24 &HS      &4.9  &45.9 &9.0e--4 &200  &1e5  &--1 &3.0 &57.5 &59.7    \\  
4C 03.24 &HS      &4.9  &45.9 &9.0e--4 &200  &1e5  &--1 &3.0 &57.5 &59.7    \\  
%         &interm. &9.7  &45.8 &1.2e--4 &300  &3e4  &--1 &3.4 &57.6 &58.8    \\  
%         &lobe    &56.5 &45.7 &1.0e--5 &900  &7e4  &1.6 &2.33 &58.9 &59.1   \\  
         &lobe    &68.2 &45.5 &6.0e--6 &900  &7e4  &1.3 &2.3  &59.0 &58.7   \\  
\hline
\hline 
\end{tabular}
\vskip 0.4 true cm
\caption{
Adopted parameters for the hotspot and lobe models shown in Fig.~\ref{4c41_sed_fig} and Fig.~\ref{4c03_sed_fig}.
Col. [1]: name; 
Col. [2]: component (HS = hotspot); 
Col. [3]: size in kpc;
Col. [4]: logarithm of the power injected in relativistic electrons in erg s$^{-1}$;
Col. [5]: magnetic field in G;
Col. [6] and Col. [7]: break and maximum Lorenz factor of the injected electron distribution;
Col. [8] and Col. [9]: slopes of the injected electron distribution; 
Col. [10]: logarithm of the total energy in relativistic electrons, in erg;
Col. [11]: logarithm of the total energy in magnetic field, in erg.
The values of the powers and the energetics refer to {\it one} jet and {\it one} hotspot and lobe,
while the lobe flux shown in the figures corresponds to {\it two} hotspots and lobes.
}
\label{paralobe_table}
\end{table*}
% --------------------------------------

\section{SED Modelling}
\label{modelsed}

Following Ghisellini et. al (2015), and references therein, we model the broadband SEDs of 4C~41.17
and 4C~03.24 as arising from the super-imposition of several components: i) an accretion disk plus
absorbing torus (and/or absorbing dust within the host galaxy); ii) a (misaligned), compact,
relativistic jet; iii) two, relatively compact hotspots (which are assumed at rest),
corresponding to the jet termination shocks; and iv) two extended lobes. The purpose of the SED
modelling is to determine the radiation mechanism for the observed flux in different bands, as
well as the origin of the emission in case that decomposing into point and extended components is
possible. 

\subsection{Disk/torus emission and black hole mass estimates}\label{modelsed:disk}

We assume a standard, geometrically thin, optically thick accretion disk \citep{shakura+1973}.
The disk is surrounded by an absorbing torus, which intercepts the disk thermal emission and re-emits in the IR band.

%Interpreting the compact radio flux of 4C 41.17 as coming from the inner
%relativistic jet, implies to assume a relatively small viewing angle. 
%We adopt 15$^\circ$.
For 4C~41.17, the broadband SED shape indicates some level of absorption occurring in the optical band. 
We start by adopting a viewing angle ($\theta_{\rm v}$) that is offset by $45^\circ$ with respect to the jet axis -- which, in turn, coincides with the orthogonal to the inner disk.
If the opening angle of the torus with respect to the jet axis is smaller than the viewing angle ($\theta_{\rm torus} < \theta_{\rm v}$), 
then the accretion disk emission as seen by the observer is absorbed\footnote{In this case, the "core" 
radio emission reported by \cite{carilli+1994} cannot possibly be produced by the relativistic jet. However, the disk emission could also be absorbed by dust belts located at relatively large distance from the black hole (and yet still within the host galaxy). In this case, $\theta_{\rm v}$ could be 
smaller, and the jet emission could indeed contribute to the
"radio core" identified by \cite{carilli+1994}.}. Regardless, for the inferred disk luminosity, $L_{\rm d}$, not to exceed the Eddington limit, $L_{\rm Edd}$, the black hole mass, $M$, must exceed $2\times 10^9M_\odot$.  In the following, we shall adopt $M\sim 5\times 10^{9} M_\odot$,
corresponding to $L_{\rm d}/L_{\rm Edd}\simeq 0.4$. Although the exact
value of $M$ is uncertain, this is not critical for the purpose of the SED modelling. \\

For 4C~03.24, the ratio between the observed IR and optical/UV emission
is more typical of unabsorbed quasars; hence, $\theta_{\rm torus}$ likely exceeds $\theta_{\rm v}$, and the disk emission
is not absorbed. Nevertheless, the highest frequency optical
flux (i.e., rest frame UV, indicated the vertical line labelled "Ly$\alpha$"
in Fig. \ref{4c03_sed_fig}) is likely absorbed by intervening
Ly$\alpha$ clouds.
The Ly$\alpha$ flux for this system corresponds to a Ly$\alpha$ luminosity of
$\sim 2\times 10^{44}$ erg s$^{-1}$ \citep{roettgering+1997}; following \citet{francis+1991} and \citet{vandenberk+2001},
this corresponds to a broad line luminosity $L_{\rm BLR}\sim  10^{45}$ erg s$^{-1}$.
Further assuming that the broad lines re-emit 10\% of the disk flux, 
this yields $L_{\rm d}\sim 10^{46}$ erg s$^{-1}$.
Adopting this value, the best agreement with the optical data is found for a black hole mass
$M=6\times 10^8 M_\odot$ (corresponding to $L_{\rm d}/L_{\rm Edd}=0.15$; lower disk luminosities would imply a larger black hole mass). 
%As indicated by the vertical line in Fig. \ref{4c03_sed_fig}, the
%highest frequency optical point lies beyond the rest-frame Ly$\alpha$
%limit, suggesting some level of absorption by intervening material. 

\subsection{Jet emission}\label{modelsed:jet}

We assume that the total jet power -- kinetic plus magnetic -- is
dissipated mostly in one zone (which, in turn, is responsible for producing the bulk of the 
radiation that is observed for very low viewing angle systems, such as blazars). This is not necessarily the case in radio galaxies, 
which are viewed at relatively large angles. For those, emission arising from the inner jet can still be detected
if the viewing angle, $\theta_{\rm v}$, is not very large and/or the jet is ``structured",
i.e., there is a gradient in bulk velocity and Lorentz
factor, either as a function of the radial distance
\citep[accelerating or decelerating jets, e.g.,][]{georganopoulos+2003}, 
or the angular distance from the jet axis
\citep[i.e., a fast spine surrounded by a slower layer,][]{ghisellini+2005}.
For simplicity, we assume here that the jet is not structured, and is observed 
at an intermediate $\theta_{\rm v}$. We set $\theta_{\rm v}=45^\circ$ for
4C~41.17, and $\theta_{\rm v}=17^\circ$ for 4C~03.17, to account for the bright  
core \xray\ emission (although, if this \xray\ component were instead produced by a somewhat less beamed 
portion of the jet, this would allow for somewhat larger values of $\theta_{\rm v}$).

%The latter source presents a relatively steep radio emission 
%coming from the radio core.
%We interpret this emission as coming not from the jet (that should
%produce a flat radio spectrum, even if de--beamed), but from a
%structure of size intermediate from the hotspot and the lobe.

The emission region is assumed to be spherical with a radius of $R=\psi R_{\rm diss}$, defined as the distance from the black hole,  
and $\psi$ is the semi-aperture angle of the conical jet (we assume $\psi=0.1$~rad).
The region moves with velocity $\beta c$, 
corresponding to a bulk Lorentz factor $\Gamma$.
Throughout, relativistic electrons are injected at a rate $Q(\gamma)$, embedded in a tangled magnetic field $B$, and have 
total power $P^\prime_{\rm e}$ (measured in the co-moving frame). 
%
% \begin{equation}
% P^\prime_{\rm e} = V m_{\rm e} c^2 \int_{\gamma_{\rm min}}^{\gamma_{\rm max}} Q(\gamma)\gamma d\gamma$,
% \end{equation}
%
% where $V$ is the volume and $\gamma_{\rm min}$ and $\gamma_{\rm max}$ are the
% minimum and maximum injection energies. 
The electron energy distribution, $Q(\gamma)$, is modelled as a double power law, with slopes $s_1$ and $s_2$ below and
above the break energy, $\gamma_{\rm b}$, respectively: 
\begin{equation}
Q(\gamma)  \, = \, Q_0\, { (\gamma/\gamma_{\rm b})^{-s_1} \over 1+
(\gamma/\gamma_{\rm b})^{-s_1+s_2} } \quad {\rm [cm^{-3} s^{-1}]}.
\label{qgamma}
\end{equation}
$Q(\gamma)$ extends between $\gamma_{\rm min}$
(assumed to be $\sim$1) and $\gamma_{\max}$.
Since we always assume that $s_1\leq 1$, $s_2>2$, 
the exact value of $\gamma_{\rm max}$ is not critical.
The particle density distribution, $N(\gamma)$ [cm$^{-3}$],
is determined by solving the continuity equation at a time corresponding to the emission region 
light-crossing time, taking into account radiative cooling as well as pair production (albeit the latter is not relevant in our cases).
Electrons radiatively cool by synchrotron and inverse Compton processes,
with seed photons provided by the accretion disk, broad line region (BLR), IR torus, and CMB. 
The jet carries power in the form of Poynting flux
$P_{\rm B}$, along with bulk motion of electrons, $P_{\rm e}$,
and protons, $P_{\rm p}$. 
In addition, we account for the power $P_{\rm r}$ spent by the jet
to produce the observed radiation. The corresponding parameters for the jet models
are listed in Table~\ref{para_table}.

We notice that, unlike for blazars, the contribution from the compact jet to the observed SED is going to be entirely negligible for 4C~41.17 and 4C~03.24. Nevertheless, the compact jet remains a key ingredient in the model, in that its kinetic jet power is directly related to the power injected in the lobes and hotspots in the form of magnetic field and relativistic electrons. Even though virtually none of the jet parameters can be directly constrained by modelling the systems' SEDs, its power is indirectly constrained by the observed optical emission via the known relation between jet power and accretion disk luminosity \citep{ghisellini+2014}. Thus, although the jet parameters listed in Table~\ref{para_table} can not be thought of as "best-fitting" parameters, they nevertheless provide us with a consistency check for the model, in that they need to not only be consistent with the measured upper limits (both to the radio and the optical emission, via the jet power relation to the disk luminosity), but also power the lobes and hotspots in such a way that the measured synchrotron and Compton luminosities satisfy the observational constraints. 

For 4C~03.24, the observed radio emission at the arcsec scale has a relatively steep spectrum
and cannot be accounted for the compact jet, whose radio spectrum is flat, nor by the hotspots,
that are of larger size.
We are then obliged to consider another emission component, that we
associate to the large scale jet, labelled in Fig. \ref{4c03_sed_fig} and in Table \ref{para_table}
as ``kpc jet".

%We  assume that  the BLR reprocesses 10\% of $L_{\rm d}$, 
%lies at a distance $R_{\rm BLR} = 10^{17}L^{1/2}_{\rm d, 45}$ cm. 
%The torus, at $R=2.5\times 10^{18} L^{1/2}_{\rm d, 45}$ cm, intercepts 
%and re--emits a fraction ($\sim$20--40\%) of $L_{\rm d}$ in the infrared band.

%The different forms of power carried by the jet can be expressed as energy fluxes as
%
%\begin{equation}
%P_{\rm i} \, =\, \pi R^2 \Gamma^2\beta c\, U^\prime_{\rm i}.
%\end{equation}
%
%where $U'_{\rm i}$ represents the energy density (in the jet comoving frame) in radiation,
%$U^\prime_{\rm r}$, corresponding to the radiative jet power $P_{\rm r}$, and in magnetic field, 
%$U^\prime_{\rm B}=B^{\prime \, 2}/(8\pi)$, corresponding to the jet Poynting flux $P_{\rm B}$.  
%Finally, we need to consider the kinetic energy density of the electrons, 
%$U^\prime_{\rm e} = m_{\rm e}c^2\, \int N(\gamma)\gamma d\gamma$, corresponding to 
%the power carried by the electrons, 
%$P_{\rm e}$, and protons, $P_{\rm p}$ (we are assuming to have one cold proton -- 
%i.e. whose kinetic energy is due to bulk motion only -- per emitting electron). 
%The total jet power, $P_{\rm jet}$, is the sum of all these components.

\subsection{Hotspot emission}\label{modelsed:hotspot}

The jet deposits its power into the
hotspots through a termination shock, which in turn energizes the more extended lobes. 
To model the radiation produced in the hotspots 
we follow the prescriptions detailed in \citet{ghisellini+2015}.
The particle injection function obeys Eq. \ref{qgamma}, with different
parameters from the jet's.
The hotspot sizes are assumed to be between 3--5 kpc, similarly to what is observed in nearby objects (such as Cygnus A, see \citealt{wilson+2000}). The amount of injected power, combined with the size, constrain the the magnetic field and particle energy density (once the particle distribution is self-consistently derived by solving the continuity equation). 
Under these assumptions, we find that the hotspots in 4C~41.17 and 4C~03.24 have magnetic fields
larger than the equipartition value with the electron energy.
The magnetic energy density, $U_{\rm B}$, is also found to exceed the CMB
photon energy density, $U_{\rm CMB}$. 
For non moving sources, the magnetic field corresponding to equipartition with the CMB photon energy density is given by the expression:
\begin{equation} 
B_{\rm CMB}\, =\,5.56\times10^{-5}\,\left ({1+z\over 4.5} \right)^2 \quad {\rm G}
\label{bcmb}
\end{equation}
For both 4C~41.17 and 4C~03.24, the radio flux in both sources is dominated
by the hotspots. As a consequence, the {\it radiative cooling is dominated by synchrotron
emission.}

\subsection{Lobe emission}\label{modelsed:lobe}

For take each lobe to be a sphere that is homogeneously filled with a magnetic field of corresponding coherence length $\lambda=10$ kpc
\citep[e.g.,][]{carilli+2002,celotti+2004}. 
The total injected power, $P_{\rm e, lobe}$, follows distribution given 
by Eq. \ref{qgamma}, with different yet parameters from the
jets' and the hotspots'.
At variance with the hotspots, we assume that, within the lobes, 
the magnetic field energy is in equipartition with the electron
energy, enabling us to infer the magnetic field value (unlike for the hotspots, where the size sets the balance between magnetic and particle energy density, the lobe sizes are basically unconstrained, forcing us to assume equipartition to derive them in turn; this yields lobe sizes of the order of $50$ kpc\footnote{Notice that, while the parameters in Table~\ref{paralobe_table} are given for a single lobe/hotspot,
fluxes plotted Fig. \ref{4c41_sed_fig} and Fig. \ref{4c03_sed_fig} refer to the cumulative fluxes from each pair of lobes/hotspots.}).
For both systems, $B_{\rm lobe}<B_{\rm CMB}$ (Eq. \ref{bcmb}), implying that 
the {\it synchrotron radio emission from the lobes is quenched, whereas \xray\ emission by the inverse Compton scattering dominates the radiative power output}.

\section{Discussion and Conclusions}

In this work, we constructed and modelled the SEDs of two high-$z$ ($z>$3.5)
radio galaxies to investigate the radiation mechanism of the extended emission
in GHz radio and \xray\ bands. One of our conclusions is that the synchrotron radio
emission from the lobes is quenched by the CMB photons which are upscattered
into the \xray\ band. 
One of the key challenges faced by CMB quenching as a viable mechanism to account for the deficit of high redshift, radio loud AGNs is provided by the fact that, for these systems, the measured 
\lx/\lr\ fails to increase with redshift as $(1+z)^4$ over a fairly broad redshift range \citep{smail+2012, smail+2013}. 
As a possible explanation, these works suggest that {\it IR photons from the host galaxy} could provide a sizable -- possibly dominant -- fraction of the seed photons available for Comptonization into the \xray\ band, thereby suppressing the strong redshift dependence that would be expected if CMB photons were solely responsible for seeding the IC process. 
If our modelling results are correct, however, this is unlikely to work above $z\simeq 2$. Specifically, the ratio between the IR and CMB photon energy densities depends on the system IR luminosity, $L_{\rm IR}$, and lobe size, $d$, as follows: 
\begin{equation}
{U_{\rm IR} \over U_{\rm CMB}} \, =\, {L_{\rm IR} \over 4\pi d^2 c U_{\rm CMB} } 
\, \sim \, 70 \, {L_{\rm IR, 46} \over d_{\rm 10\, kpc}^2 (1+z)^4 }
\end{equation}
Given the inferred values of $d$ and $L_{\rm IR}$, the IR and CMB energy density contributions are still comparable at $z\sim 2$, above which $U_{\rm CMB}$ starts to prevail. 

Furthermore, the main conclusion from our modelling is that the hotspots of 4C~41.17 and 4C~03.24 would be magnetically dominated. If so, {independent 
from the ratio $U_{\rm IR}/U_{\rm CMB}$}, most of the \xray\ signal ought to be produced within the lobes, with only a negligible contribution from the hotspots. Conversely, any extended radio emission likely originates from the hotspots themselves, rather than the lobes. 
As a consequence, even though the {\it intrinsic \lx/\lr\ within the lobes
should still increase as $(1+z)^4$, the face-value, {\it measured} ratio, does not display any redshift dependence. }
%This also washes out any expected, redshift-dependent increase of the observed X-ray to radio luminosity ratios for these radio galaxies, as the expected  $(1+z)^4$ dependence would only be observable if the lobes were dominating the emission in both bands. 

In summary, although not aiming to completely solve the problem of
  missing high-redshift radio-loud AGNs, our work supports the viability
  of CMB quenching as an effective mechanism to significantly dim the
  diffuse radio emission from intrinsically jetted AGNs at high redshifts. For
  systems where the extended radio emission is dominated by the lobes instead of
  the hotspots, CMB quenching would cast them into the radio-quiet regime.
However, while CMB quenching remains firmly in play for high redshift radio galaxies, the limited
sample under consideration in this work does not enable us to confirm to what extent this mechanism
might be {\it entirely} sufficient to explain the apparent deficit of high-$z$ radio-loud AGNs, or
whether intrinsic obscuration of the central engine might be needed in addition \citep{ghisellini+2016}.

We close by acknowledging that the modelling of radio emission in this work --
specifically the relative contribution of the lobes vs. hotspots -- has some
level of degeneracy, in that the hotspots are not spatially resolved at these
redshifts. Upcoming low-frequency radio observations ($\sim$MHz), with the LOw
Frequency Array \citep[LOFAR;][]{vanhaarlem+2013}, should be able to isolate the
contribution to the diffuse radio signal from the lobes alone \cite{ghisellini+2015}. 

% -----------------------------------------------------------------------------
% Discussion
% -----------------------------------------------------------------------------

%\section{Summary \& Discussion}\label{discuss}

% -----------------------------------------------------------------------------
% Acknowledgments
% -----------------------------------------------------------------------------

\section*{Acknowledgments}
%{\bf (Acknowledgments go here.)}
G.G. and F.T. acknowledge contribution from grant PRIN--INAF--2014. This research has made use of the NASA/IPAC Extragalactic Database (NED) which is operated by the Jet Propulsion Laboratory, California Institute of Technology, under contract with the National Aeronautics and Space Administration.

% -----------------------------------------------------------------------------
% Appendix
% -----------------------------------------------------------------------------

% -----------------------------------------------------------------------------
% Bibliography
% -----------------------------------------------------------------------------

\bibliographystyle{mnras}
\bibliography{master}

\begin{thebibliography}{}
\makeatletter
\relax
\def\mn@urlcharsother{\let\do\@makeother \do\$\do\&\do\#\do\^\do\_\do\%\do\~}
\def\mn@doi{\begingroup\mn@urlcharsother \@ifnextchar [ {\mn@doi@}
  {\mn@doi@[]}}
\def\mn@doi@[#1]#2{\def\@tempa{#1}\ifx\@tempa\@empty \href
  {http://dx.doi.org/#2} {doi:#2}\else \href {http://dx.doi.org/#2} {#1}\fi
  \endgroup}
\def\mn@eprint#1#2{\mn@eprint@#1:#2::\@nil}
\def\mn@eprint@arXiv#1{\href {http://arxiv.org/abs/#1} {{\tt arXiv:#1}}}
\def\mn@eprint@dblp#1{\href {http://dblp.uni-trier.de/rec/bibtex/#1.xml}
  {dblp:#1}}
\def\mn@eprint@#1:#2:#3:#4\@nil{\def\@tempa {#1}\def\@tempb {#2}\def\@tempc
  {#3}\ifx \@tempc \@empty \let \@tempc \@tempb \let \@tempb \@tempa \fi \ifx
  \@tempb \@empty \def\@tempb {arXiv}\fi \@ifundefined
  {mn@eprint@\@tempb}{\@tempb:\@tempc}{\expandafter \expandafter \csname
  mn@eprint@\@tempb\endcsname \expandafter{\@tempc}}}

\bibitem[\protect\citeauthoryear{{Ahn} et~al.,}{{Ahn} et~al.}{2012}]{ahn+2012}
{Ahn} C.~P.,  et~al., 2012, \mn@doi [\apjs] {10.1088/0067-0049/203/2/21}, \href
  {http://adsabs.harvard.edu/abs/2012ApJS..203...21A} {203, 21}

\bibitem[\protect\citeauthoryear{{Ajello} et~al.,}{{Ajello}
  et~al.}{2009}]{ajello+2009}
{Ajello} M.,  et~al., 2009, \mn@doi [\apj] {10.1088/0004-637X/699/1/603}, \href
  {http://adsabs.harvard.edu/abs/2009ApJ...699..603A} {699, 603}

\bibitem[\protect\citeauthoryear{{Arnaud}}{{Arnaud}}{1996}]{arnaud+1996}
{Arnaud} K.~A.,  1996, in {Jacoby} G.~H.,  {Barnes} J.,  eds,  Astronomical
  Society of the Pacific Conference Series Vol. 101, Astronomical Data Analysis
  Software and Systems V. p.~17

\bibitem[\protect\citeauthoryear{{Becker}, {White}  \& {Helfand}}{{Becker}
  et~al.}{1995}]{becker+1995}
{Becker} R.~H.,  {White} R.~L.,   {Helfand} D.~J.,  1995, \mn@doi [\apj]
  {10.1086/176166}, \href {http://adsabs.harvard.edu/abs/1995ApJ...450..559B}
  {450, 559}

\bibitem[\protect\citeauthoryear{{Carilli} \& {Taylor}}{{Carilli} \&
  {Taylor}}{2002}]{carilli+2002}
{Carilli} C.~L.,  {Taylor} G.~B.,  2002, \mn@doi [\araa]
  {10.1146/annurev.astro.40.060401.093852}, \href
  {http://adsabs.harvard.edu/abs/2002ARA%26A..40..319C} {40, 319}

\bibitem[\protect\citeauthoryear{{Carilli}, {Owen}  \& {Harris}}{{Carilli}
  et~al.}{1994}]{carilli+1994}
{Carilli} C.~L.,  {Owen} F.~N.,   {Harris} D.~E.,  1994, \mn@doi [\aj]
  {10.1086/116870}, \href {http://adsabs.harvard.edu/abs/1994AJ....107..480C}
  {107, 480}

\bibitem[\protect\citeauthoryear{{Carter}, {Karovska}, {Jerius}, {Glotfelty}
  \& {Beikman}}{{Carter} et~al.}{2003}]{carter+2003}
{Carter} C.,  {Karovska} M.,  {Jerius} D.,  {Glotfelty} K.,   {Beikman} S.,
  2003, in {Payne} H.~E.,  {Jedrzejewski} R.~I.,   {Hook} R.~N.,  eds,
  Astronomical Society of the Pacific Conference Series Vol. 295, Astronomical
  Data Analysis Software and Systems XII. p.~477

\bibitem[\protect\citeauthoryear{{Cash}}{{Cash}}{1979}]{cash+1979}
{Cash} W.,  1979, \mn@doi [\apj] {10.1086/156922}, \href
  {http://adsabs.harvard.edu/abs/1979ApJ...228..939C} {228, 939}

\bibitem[\protect\citeauthoryear{{Celotti} \& {Fabian}}{{Celotti} \&
  {Fabian}}{2004}]{celotti+2004}
{Celotti} A.,  {Fabian} A.~C.,  2004, \mn@doi [\mnras]
  {10.1111/j.1365-2966.2004.08085.x}, \href
  {http://adsabs.harvard.edu/abs/2004MNRAS.353..523C} {353, 523}

\bibitem[\protect\citeauthoryear{{Chambers}, {Miley}  \& {van
  Breugel}}{{Chambers} et~al.}{1990}]{chambers+1990}
{Chambers} K.~C.,  {Miley} G.~K.,   {van Breugel} W.~J.~M.,  1990, \mn@doi
  [\apj] {10.1086/169316}, \href
  {http://adsabs.harvard.edu/abs/1990ApJ...363...21C} {363, 21}

\bibitem[\protect\citeauthoryear{{Cheung}, {Stawarz}  \&
  {Siemiginowska}}{{Cheung} et~al.}{2006}]{cheung+2006}
{Cheung} C.~C.,  {Stawarz} {\L}.,   {Siemiginowska} A.,  2006, \mn@doi [\apj]
  {10.1086/506908}, \href {http://adsabs.harvard.edu/abs/2006ApJ...650..679C}
  {650, 679}

\bibitem[\protect\citeauthoryear{{Cheung}, {Stawarz}, {Siemiginowska},
  {Gobeille}, {Wardle}, {Harris}  \& {Schwartz}}{{Cheung}
  et~al.}{2012}]{cheung+2012}
{Cheung} C.~C.,  {Stawarz} {\L}.,  {Siemiginowska} A.,  {Gobeille} D.,
  {Wardle} J.~F.~C.,  {Harris} D.~E.,   {Schwartz} D.~A.,  2012, \mn@doi
  [\apjl] {10.1088/2041-8205/756/1/L20}, \href
  {http://adsabs.harvard.edu/abs/2012ApJ...756L..20C} {756, L20}

\bibitem[\protect\citeauthoryear{{Croston}, {Hardcastle}, {Harris}, {Belsole},
  {Birkinshaw}  \& {Worrall}}{{Croston} et~al.}{2005}]{croston+2005}
{Croston} J.~H.,  {Hardcastle} M.~J.,  {Harris} D.~E.,  {Belsole} E.,
  {Birkinshaw} M.,   {Worrall} D.~M.,  2005, \mn@doi [\apj] {10.1086/430170},
  \href {http://adsabs.harvard.edu/abs/2005ApJ...626..733C} {626, 733}

\bibitem[\protect\citeauthoryear{{Davis} et~al.,}{{Davis}
  et~al.}{2012}]{davis+2012}
{Davis} J.~E.,  et~al., 2012, in Society of Photo-Optical Instrumentation
  Engineers (SPIE) Conference Series. p. 84431A, \mn@doi{10.1117/12.926937}

\bibitem[\protect\citeauthoryear{{Dickey} \& {Lockman}}{{Dickey} \&
  {Lockman}}{1990}]{dickey+1990}
{Dickey} J.~M.,  {Lockman} F.~J.,  1990, \mn@doi [\araa]
  {10.1146/annurev.aa.28.090190.001243}, \href
  {http://adsabs.harvard.edu/abs/1990ARA%26A..28..215D} {28, 215}

\bibitem[\protect\citeauthoryear{{Fabian}, {Walker}, {Celotti}, {Ghisellini},
  {Mocz}, {Blundell}  \& {McMahon}}{{Fabian} et~al.}{2014}]{fabian+2014}
{Fabian} A.~C.,  {Walker} S.~A.,  {Celotti} A.,  {Ghisellini} G.,  {Mocz} P.,
  {Blundell} K.~M.,   {McMahon} R.~G.,  2014, \mn@doi [\mnras]
  {10.1093/mnrasl/slu065}, \href
  {http://adsabs.harvard.edu/abs/2014MNRAS.442L..81F} {442, L81}

\bibitem[\protect\citeauthoryear{{Francis}, {Hewett}, {Foltz}, {Chaffee},
  {Weymann}  \& {Morris}}{{Francis} et~al.}{1991}]{francis+1991}
{Francis} P.~J.,  {Hewett} P.~C.,  {Foltz} C.~B.,  {Chaffee} F.~H.,  {Weymann}
  R.~J.,   {Morris} S.~L.,  1991, \mn@doi [\apj] {10.1086/170066}, \href
  {http://adsabs.harvard.edu/abs/1991ApJ...373..465F} {373, 465}

\bibitem[\protect\citeauthoryear{{Freeman}, {Doe}  \&
  {Siemiginowska}}{{Freeman} et~al.}{2001}]{freeman+2001}
{Freeman} P.,  {Doe} S.,   {Siemiginowska} A.,  2001, in {Starck} J.-L.,
  {Murtagh} F.~D.,  eds,  Society of Photo-Optical Instrumentation Engineers
  (SPIE) Conference Series Vol. 4477, Astronomical Data Analysis. pp 76--87
  (\mn@eprint {} {astro-ph/0108426}), \mn@doi{10.1117/12.447161}

\bibitem[\protect\citeauthoryear{{Freeman}, {Kashyap}, {Rosner}  \&
  {Lamb}}{{Freeman} et~al.}{2002}]{freeman+2002}
{Freeman} P.~E.,  {Kashyap} V.,  {Rosner} R.,   {Lamb} D.~Q.,  2002, \mn@doi
  [\apjs] {10.1086/324017}, \href
  {http://adsabs.harvard.edu/abs/2002ApJS..138..185F} {138, 185}

\bibitem[\protect\citeauthoryear{{Garmire}, {Bautz}, {Ford}, {Nousek}  \&
  {Ricker}}{{Garmire} et~al.}{2003}]{garmire+2003}
{Garmire} G.~P.,  {Bautz} M.~W.,  {Ford} P.~G.,  {Nousek} J.~A.,   {Ricker} Jr.
  G.~R.,  2003, in {Truemper} J.~E.,  {Tananbaum} H.~D.,  eds,  Society of
  Photo-Optical Instrumentation Engineers (SPIE) Conference Series Vol. 4851,
  X-Ray and Gamma-Ray Telescopes and Instruments for Astronomy.. pp 28--44,
  \mn@doi{10.1117/12.461599}

\bibitem[\protect\citeauthoryear{{Gehrels}}{{Gehrels}}{1986}]{gehrels+1986}
{Gehrels} N.,  1986, \mn@doi [\apj] {10.1086/164079}, \href
  {http://adsabs.harvard.edu/abs/1986ApJ...303..336G} {303, 336}

\bibitem[\protect\citeauthoryear{{Georganopoulos} \&
  {Kazanas}}{{Georganopoulos} \& {Kazanas}}{2003}]{georganopoulos+2003}
{Georganopoulos} M.,  {Kazanas} D.,  2003, \mn@doi [\apjl] {10.1086/378557},
  \href {http://adsabs.harvard.edu/abs/2003ApJ...594L..27G} {594, L27}

\bibitem[\protect\citeauthoryear{{Ghisellini} \& {Sbarrato}}{{Ghisellini} \&
  {Sbarrato}}{2016}]{ghisellini+2016}
{Ghisellini} G.,  {Sbarrato} T.,  2016, preprint, \href
  {http://adsabs.harvard.edu/abs/2016arXiv160305684G} {} (\mn@eprint {arXiv}
  {1603.05684})

\bibitem[\protect\citeauthoryear{{Ghisellini}, {Tavecchio}  \&
  {Chiaberge}}{{Ghisellini} et~al.}{2005}]{ghisellini+2005}
{Ghisellini} G.,  {Tavecchio} F.,   {Chiaberge} M.,  2005, \mn@doi [\aap]
  {10.1051/0004-6361:20041404}, \href
  {http://adsabs.harvard.edu/abs/2005A%26A...432..401G} {432, 401}

\bibitem[\protect\citeauthoryear{{Ghisellini}, {Celotti}, {Tavecchio}, {Haardt}
   \& {Sbarrato}}{{Ghisellini} et~al.}{2014}]{ghisellini+2014}
{Ghisellini} G.,  {Celotti} A.,  {Tavecchio} F.,  {Haardt} F.,   {Sbarrato} T.,
   2014, \mn@doi [\mnras] {10.1093/mnras/stt2394}, \href
  {http://adsabs.harvard.edu/abs/2014MNRAS.438.2694G} {438, 2694}

\bibitem[\protect\citeauthoryear{{Ghisellini}, {Haardt}, {Ciardi}, {Sbarrato},
  {Gallo}, {Tavecchio}  \& {Celotti}}{{Ghisellini}
  et~al.}{2015}]{ghisellini+2015}
{Ghisellini} G.,  {Haardt} F.,  {Ciardi} B.,  {Sbarrato} T.,  {Gallo} E.,
  {Tavecchio} F.,   {Celotti} A.,  2015, \mn@doi [\mnras]
  {10.1093/mnras/stv1541}, \href
  {http://adsabs.harvard.edu/abs/2015MNRAS.452.3457G} {452, 3457}

\bibitem[\protect\citeauthoryear{{Haiman}, {Quataert}  \& {Bower}}{{Haiman}
  et~al.}{2004}]{haiman+2004}
{Haiman} Z.,  {Quataert} E.,   {Bower} G.~C.,  2004, \mn@doi [\apj]
  {10.1086/422834}, \href {http://adsabs.harvard.edu/abs/2004ApJ...612..698H}
  {612, 698}

\bibitem[\protect\citeauthoryear{{Kratzer} \& {Richards}}{{Kratzer} \&
  {Richards}}{2015}]{kratzer+2015}
{Kratzer} R.~M.,  {Richards} G.~T.,  2015, \mn@doi [\aj]
  {10.1088/0004-6256/149/2/61}, \href
  {http://adsabs.harvard.edu/abs/2015AJ....149...61K} {149, 61}

\bibitem[\protect\citeauthoryear{{Krimm} et~al.,}{{Krimm}
  et~al.}{2013}]{krimm+2013}
{Krimm} H.~A.,  et~al., 2013, \mn@doi [\apjs] {10.1088/0067-0049/209/1/14},
  \href {http://adsabs.harvard.edu/abs/2013ApJS..209...14K} {209, 14}

\bibitem[\protect\citeauthoryear{{Li}, {Kastner}, {Prigozhin}, {Schulz},
  {Feigelson}  \& {Getman}}{{Li} et~al.}{2004}]{li+2004}
{Li} J.,  {Kastner} J.~H.,  {Prigozhin} G.~Y.,  {Schulz} N.~S.,  {Feigelson}
  E.~D.,   {Getman} K.~V.,  2004, \mn@doi [\apj] {10.1086/421866}, \href
  {http://adsabs.harvard.edu/abs/2004ApJ...610.1204L} {610, 1204}

\bibitem[\protect\citeauthoryear{{McGreer}, {Helfand}  \& {White}}{{McGreer}
  et~al.}{2009}]{mcgreer+2009}
{McGreer} I.~D.,  {Helfand} D.~J.,   {White} R.~L.,  2009, \mn@doi [\aj]
  {10.1088/0004-6256/138/6/1925}, \href
  {http://adsabs.harvard.edu/abs/2009AJ....138.1925M} {138, 1925}

\bibitem[\protect\citeauthoryear{{McKeough} et~al.,}{{McKeough}
  et~al.}{2016}]{mckeough+2016}
{McKeough} K.,  et~al., 2016, preprint, \href
  {http://adsabs.harvard.edu/abs/2016arXiv160903425M} {} (\mn@eprint {arXiv}
  {1609.03425})

\bibitem[\protect\citeauthoryear{{Mocz}, {Fabian}  \& {Blundell}}{{Mocz}
  et~al.}{2011}]{mocz+2011}
{Mocz} P.,  {Fabian} A.~C.,   {Blundell} K.~M.,  2011, \mn@doi [\mnras]
  {10.1111/j.1365-2966.2011.18198.x}, \href
  {http://adsabs.harvard.edu/abs/2011MNRAS.413.1107M} {413, 1107}

\bibitem[\protect\citeauthoryear{{Mocz}, {Fabian}  \& {Blundell}}{{Mocz}
  et~al.}{2013}]{mocz+2013}
{Mocz} P.,  {Fabian} A.~C.,   {Blundell} K.~M.,  2013, \mn@doi [\mnras]
  {10.1093/mnras/stt689}, \href
  {http://adsabs.harvard.edu/abs/2013MNRAS.432.3381M} {432, 3381}

\bibitem[\protect\citeauthoryear{{Morrison} \& {McCammon}}{{Morrison} \&
  {McCammon}}{1983}]{morrison+1983}
{Morrison} R.,  {McCammon} D.,  1983, \mn@doi [\apj] {10.1086/161102}, \href
  {http://adsabs.harvard.edu/abs/1983ApJ...270..119M} {270, 119}

\bibitem[\protect\citeauthoryear{{Nousek} \& {Shue}}{{Nousek} \&
  {Shue}}{1989}]{nousek+1989}
{Nousek} J.~A.,  {Shue} D.~R.,  1989, \mn@doi [\apj] {10.1086/167676}, \href
  {http://adsabs.harvard.edu/abs/1989ApJ...342.1207N} {342, 1207}

\bibitem[\protect\citeauthoryear{{Roettgering}, {van Ojik}, {Miley},
  {Chambers}, {van Breugel}  \& {de Koff}}{{Roettgering}
  et~al.}{1997}]{roettgering+1997}
{Roettgering} H.~J.~A.,  {van Ojik} R.,  {Miley} G.~K.,  {Chambers} K.~C.,
  {van Breugel} W.~J.~M.,   {de Koff} S.,  1997, \aap, \href
  {http://adsabs.harvard.edu/abs/1997A%26A...326..505R} {326, 505}

\bibitem[\protect\citeauthoryear{{Sbarrato}, {Ghisellini}, {Nardini},
  {Tagliaferri}, {Greiner}, {Rau}  \& {Schady}}{{Sbarrato}
  et~al.}{2013}]{sbarrato+2013}
{Sbarrato} T.,  {Ghisellini} G.,  {Nardini} M.,  {Tagliaferri} G.,  {Greiner}
  J.,  {Rau} A.,   {Schady} P.,  2013, \mn@doi [\mnras] {10.1093/mnras/stt882},
  \href {http://adsabs.harvard.edu/abs/2013MNRAS.433.2182S} {433, 2182}

\bibitem[\protect\citeauthoryear{{Sbarrato}, {Ghisellini}, {Tagliaferri},
  {Foschini}, {Nardini}, {Tavecchio}  \& {Gehrels}}{{Sbarrato}
  et~al.}{2015}]{sbarrato+2015}
{Sbarrato} T.,  {Ghisellini} G.,  {Tagliaferri} G.,  {Foschini} L.,  {Nardini}
  M.,  {Tavecchio} F.,   {Gehrels} N.,  2015, \mn@doi [\mnras]
  {10.1093/mnras/stu2269}, \href
  {http://adsabs.harvard.edu/abs/2015MNRAS.446.2483S} {446, 2483}

\bibitem[\protect\citeauthoryear{{Scharf}, {Smail}, {Ivison}, {Bower}, {van
  Breugel}  \& {Reuland}}{{Scharf} et~al.}{2003}]{scharf+2003}
{Scharf} C.,  {Smail} I.,  {Ivison} R.,  {Bower} R.,  {van Breugel} W.,
  {Reuland} M.,  2003, \mn@doi [\apj] {10.1086/377531}, \href
  {http://adsabs.harvard.edu/abs/2003ApJ...596..105S} {596, 105}

\bibitem[\protect\citeauthoryear{{Shakura} \& {Sunyaev}}{{Shakura} \&
  {Sunyaev}}{1973}]{shakura+1973}
{Shakura} N.~I.,  {Sunyaev} R.~A.,  1973, \aap, \href
  {http://adsabs.harvard.edu/abs/1973A%26A....24..337S} {24, 337}

\bibitem[\protect\citeauthoryear{{Siemiginowska}, {Smith}, {Aldcroft},
  {Schwartz}, {Paerels}  \& {Petric}}{{Siemiginowska}
  et~al.}{2003}]{siemiginowska+2003}
{Siemiginowska} A.,  {Smith} R.~K.,  {Aldcroft} T.~L.,  {Schwartz} D.~A.,
  {Paerels} F.,   {Petric} A.~O.,  2003, \mn@doi [\apjl] {10.1086/380497},
  \href {http://adsabs.harvard.edu/abs/2003ApJ...598L..15S} {598, L15}

\bibitem[\protect\citeauthoryear{{Skrutskie} et~al.,}{{Skrutskie}
  et~al.}{2006}]{skrutskie+2006}
{Skrutskie} M.~F.,  et~al., 2006, \mn@doi [\aj] {10.1086/498708}, \href
  {http://adsabs.harvard.edu/abs/2006AJ....131.1163S} {131, 1163}

\bibitem[\protect\citeauthoryear{{Smail} \& {Blundell}}{{Smail} \&
  {Blundell}}{2013}]{smail+2013}
{Smail} I.,  {Blundell} K.~M.,  2013, \mn@doi [\mnras] {10.1093/mnras/stt1240},
  \href {http://adsabs.harvard.edu/abs/2013MNRAS.434.3246S} {434, 3246}

\bibitem[\protect\citeauthoryear{{Smail} et~al.,}{{Smail}
  et~al.}{2009}]{smail+2009}
{Smail} I.,  et~al., 2009, \mn@doi [\apjl] {10.1088/0004-637X/702/2/L114},
  \href {http://adsabs.harvard.edu/abs/2009ApJ...702L.114S} {702, L114}

\bibitem[\protect\citeauthoryear{{Smail}, {Blundell}, {Lehmer}  \&
  {Alexander}}{{Smail} et~al.}{2012}]{smail+2012}
{Smail} I.,  {Blundell} K.~M.,  {Lehmer} B.~D.,   {Alexander} D.~M.,  2012,
  \mn@doi [\apj] {10.1088/0004-637X/760/2/132}, \href
  {http://adsabs.harvard.edu/abs/2012ApJ...760..132S} {760, 132}

\bibitem[\protect\citeauthoryear{{Vanden Berk} et~al.,}{{Vanden Berk}
  et~al.}{2001}]{vandenberk+2001}
{Vanden Berk} D.~E.,  et~al., 2001, \mn@doi [\aj] {10.1086/321167}, \href
  {http://adsabs.harvard.edu/abs/2001AJ....122..549V} {122, 549}

\bibitem[\protect\citeauthoryear{{Volonteri}, {Haardt}, {Ghisellini}  \& {Della
  Ceca}}{{Volonteri} et~al.}{2011}]{volonteri+2011}
{Volonteri} M.,  {Haardt} F.,  {Ghisellini} G.,   {Della Ceca} R.,  2011,
  \mn@doi [\mnras] {10.1111/j.1365-2966.2011.19024.x}, \href
  {http://adsabs.harvard.edu/abs/2011MNRAS.416..216V} {416, 216}

\bibitem[\protect\citeauthoryear{{Wilson}, {Young}  \& {Shopbell}}{{Wilson}
  et~al.}{2000}]{wilson+2000}
{Wilson} A.~S.,  {Young} A.~J.,   {Shopbell} P.~L.,  2000, \mn@doi [\apjl]
  {10.1086/317293}, \href {http://adsabs.harvard.edu/abs/2000ApJ...544L..27W}
  {544, L27}

\bibitem[\protect\citeauthoryear{{Wu}, {Brandt}, {Miller}, {Garmire},
  {Schneider}  \& {Vignali}}{{Wu} et~al.}{2013}]{wu+2013}
{Wu} J.,  {Brandt} W.~N.,  {Miller} B.~P.,  {Garmire} G.~P.,  {Schneider}
  D.~P.,   {Vignali} C.,  2013, \mn@doi [\apj] {10.1088/0004-637X/763/2/109},
  \href {http://adsabs.harvard.edu/abs/2013ApJ...763..109W} {763, 109}

\bibitem[\protect\citeauthoryear{{York} et~al.,}{{York}
  et~al.}{2000}]{york+2000}
{York} D.~G.,  et~al., 2000, \mn@doi [\aj] {10.1086/301513}, \href
  {http://adsabs.harvard.edu/abs/2000AJ....120.1579Y} {120, 1579}

\bibitem[\protect\citeauthoryear{{Yuan}, {Fabian}, {Worsley}  \&
  {McMahon}}{{Yuan} et~al.}{2006}]{yuan+2006}
{Yuan} W.,  {Fabian} A.~C.,  {Worsley} M.~A.,   {McMahon} R.~G.,  2006, \mn@doi
  [\mnras] {10.1111/j.1365-2966.2006.10175.x}, \href
  {http://adsabs.harvard.edu/abs/2006MNRAS.368..985Y} {368, 985}

\bibitem[\protect\citeauthoryear{{van Haarlem} et~al.,}{{van Haarlem}
  et~al.}{2013}]{vanhaarlem+2013}
{van Haarlem} M.~P.,  et~al., 2013, \mn@doi [\aap]
  {10.1051/0004-6361/201220873}, \href
  {http://adsabs.harvard.edu/abs/2013A%26A...556A...2V} {556, A2}

\bibitem[\protect\citeauthoryear{{van Ojik}, {Roettgering}, {Carilli}, {Miley},
  {Bremer}  \& {Macchetto}}{{van Ojik} et~al.}{1996}]{vanojik+1996}
{van Ojik} R.,  {Roettgering} H.~J.~A.,  {Carilli} C.~L.,  {Miley} G.~K.,
  {Bremer} M.~N.,   {Macchetto} F.,  1996, \aap, \href
  {http://adsabs.harvard.edu/abs/1996A%26A...313...25V} {313, 25}

\makeatother
\end{thebibliography}

% -----------------------------------------------------------------------------
% Tables
% -----------------------------------------------------------------------------
%\clearpage

%\clearpage

% -----------------------------------------------------------------------------
% Figures
% -----------------------------------------------------------------------------

\clearpage
% Fig. 1: The Chandra ACIS image of 4C 41.17. 

\clearpage
% Fig. 2: The Chandra ACIS image of 4C 03.24. 

\clearpage
% Fig. 3: The Chandra ACIS image of other sources. 

\clearpage
\newpage

% Fig. 4: The radial photon density profile of 4C 41.17.

\clearpage
\newpage

% Fig. 5: The radial photon density profile of 4C 03.24.

\clearpage
% Fig. 6: The radial photon density profile of 4C 19.71, 4C 60.07, and TN J1338.

\clearpage
% Fig. 7: The SED of 4C 41.17. 

\clearpage
% Fig. 8: The SED of 4C 03.24. 

\end{document}